\documentclass[lettersize,journal]{IEEEtran}
\usepackage{soul}
\usepackage[utf8]{inputenc}
\usepackage{amsthm}

\newtheorem{definition}{Definition}
\usepackage{epsfig, subfigure, amsmath, amssymb}
\usepackage{textcomp}
\usepackage{url}
\usepackage{wrapfig,lipsum,booktabs}
\usepackage{multirow}
\usepackage{amsmath,amssymb,amsfonts}
\usepackage[ruled,vlined,linesnumbered]{algorithm2e}
\usepackage{algorithmic}
\usepackage{graphicx}
\usepackage{textcomp}
\usepackage{multirow}
\usepackage{xcolor}
\usepackage{amsmath}
\usepackage{paralist}
\begin{document}

\title{MISGUIDE: Security-Aware Attack Analytics for Smart Grid Load Frequency Control}

\author{Nur~Imtiazul~Haque,~\IEEEmembership{Member,~IEEE,}
        Prabin~Mali,~\IEEEmembership{Member,~IEEE,}
        Mohammad~Zakaria~Haider,~\IEEEmembership{Member,~IEEE,}
        Mohammad~Ashiqur~Rahman,~\IEEEmembership{Senior Member,~IEEE,}
        Sumit~Paudyal,~\IEEEmembership{Senior Member,~IEEE}}

\definecolor{emerald}{rgb}{0.00784313725 ,0.54117647058,0.05882352941}

\newcommand{\nur}[1]{\textit{\color{emerald} #1}}

\newcommand{\revise}[1]{\textit{\color{red} #1}}

\newcommand{\remark}[1]{\textit{\color{red} #1}}
\newcommand{\prabin}[1]{\textit{\color{blue} #1}}

\newcommand{\framework}{MISGUIDE}

\maketitle

\begin{abstract}
Incorporating advanced information and communication technologies into smart grids (SGs) offers substantial operational benefits while increasing vulnerability to cyber threats like false data injection (FDI) attacks. Current SG attack analysis tools predominantly employ formal methods or adversarial machine learning (ML) techniques with rule-based bad data detectors to analyze the attack space. However, these attack analytics either generate simplistic attack vectors detectable by the ML-based anomaly detection models (ADMs) or fail to identify critical attack vectors from complex controller dynamics in a feasible time. This paper introduces MISGUIDE, a novel defense-aware attack analytics designed to extract verifiable multi-time slot-based FDI attack vectors from complex SG load frequency control dynamics and ADMs, utilizing the Gurobi optimizer. MISGUIDE can identify optimal (maliciously triggering under/over frequency relays in minimal time) and stealthy attack vectors. Using real-world load data, we validate the MISGUIDE-identified attack vectors through real-time hardware-in-the-loop (OPALRT) simulations of the IEEE 39-bus system.
\end{abstract}


%%%%%%%%%%%%%%%%%%%%%%%%%%%%%
\vspace{-10pt}
%%%%%%%%%%%%%%%%%%%%%%%%%%%%%%%%%%%%%%%%%%%%%
%%%%%%%%%%%%%%%%%%%%%%%%%%%%%%%%%%%%%%%%%%%%%
\section{Introduction}
With the integration of advanced information and communication technologies, traditional power grids are evolving into smart grids (SGs)~\cite{yu2016sg}. Although this integration brings numerous operational benefits, such as improved efficiency for demand response, it also enlarges the attack surface of power grids to cyber threats due to their high dependence on the vulnerable communication infrastructure~\cite{bose2010smart}. According to the 2022 threat intelligence index report, 10.7\% of all cyberattacks targeted the energy sector, making it the fourth most attacked industry~\cite{gregory2023securityintelligence}. One significant category of cyberattacks that has received extensive attention in recent research is the false data injection (FDI) attack~\cite{li2017cybersecurity, liang2015, bobba2010detecting, liu2011false}. Several real-world incidents have been reported that caused damage to the SGs due to FDI attacks. For example, in 2015, an FDI attack compromised distribution grids in Ukraine, where hackers infiltrated the communication network, causing widespread power outages and affecting over 200,000 customers for several hours~\cite{liang2015}. Other notable real-world attacks include Stuxnet~\cite{karnouskos2011stuxnet} and Dragonfly~\cite{Dragonfly}, where attackers gained extensive knowledge of the targeted systems and complete access to real-time data in the control centers of critical SG infrastructures.

\begin{figure}[!t]
\centering
\includegraphics[width=0.99\columnwidth]{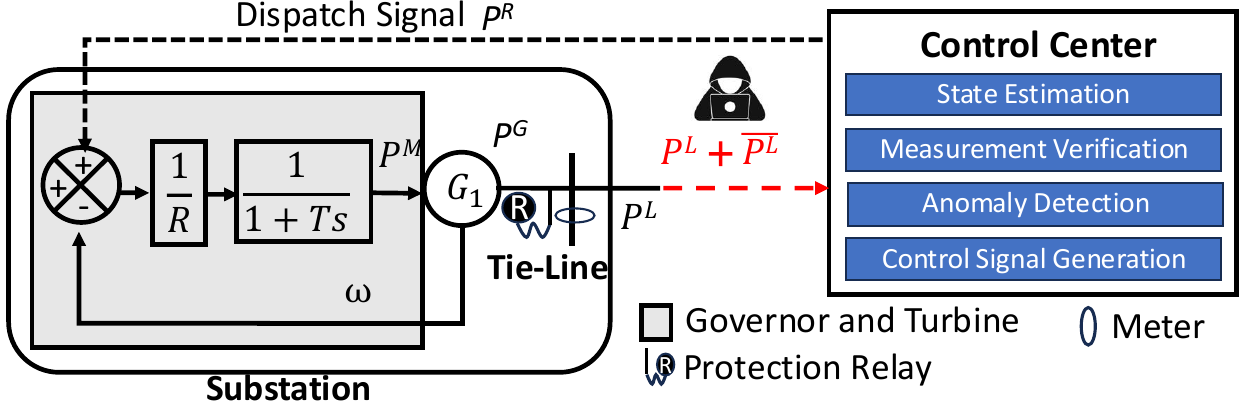}
\caption{Demonstration of generators dispatching process and possible point of attacks in SG.}
\label{fig:lfc}
\vspace{-18pt}
\end{figure}

Grid frequency is a crucial indicator of the stable operation of an SG. The primary controllers of the grid generators and a central load frequency controller (LFC) maintain the grid frequency (60 Hz in the US). Deviation from the nominal/standard frequency can result in undesirable effects on the SG, like degradation of load performance, equipment damage, or interference in the protection scheme, which may ultimately lead to grid instability. In response to disturbances, e.g., increased load demand in power grids, the primary frequency response utilizes the automatic decentralized control actions of generators' active power output to regulate the grid frequency instantaneously. Although the primary controller ensures synchronization of the generator frequency, it cannot restore the system frequency to its nominal value, as the grid's load demand continuously fluctuates. Hence, a centralized LFC involves re-dispatching the generators to maintain the grid frequency within the desired range as depicted in Figure~\ref{fig:lfc}. The SG uses several protection relays to handle underfrequency (UF) and overfrequency (OF) instances. These relays are essential in maintaining a balance between load and generation by disconnecting generators/loads at pre-defined locations as necessary~\cite{rocof2020}. For example, the OF relays trip generators during excessive frequency excursions to protect the system from frequency instability, thus protecting synchronous generators from causing overheating, mechanical stress, and voltage instability and contributing to broader system instability issues. While most protection relays operate based on local measurements, the occurrence of an FDI attack in the closed-loop control of the dispatching process can cause extreme frequency variations, ultimately leading to false relay operations (FRO), resulting in erroneous tripping of UF or OF relays and far-reaching consequences for the entire system. In an FDI attack, attackers can manipulate and send erroneous load and frequency measurements to the SG communication system as shown in Figure~\ref{fig:lfc}, deceiving the LFC into making the wrong reference setpoint of the generators. The figure illustrates the communication between a single bus or substation and the controller.
%%%%%%%%%% Figure: LFC and Attack Points %%%%%%%%%%

In response to these challenges, modern grid systems adopt different machine learning (ML)-based anomaly detection models (ADM) to detect abnormal events in safety-critical SG systems. Hence, to understand the actual robustness of the system, it is imperative to analyze the attack space of the system in the presence of the ADM. Traditional approaches, including formal analysis and ML techniques like adversarial ML and reinforcement learning (RL), have shown promise in attack vector identification. Adversarial ML and RL-based techniques are efficient and fast, yet they often fail to extract attacks that can evade ML-based ADMs. The primary limitation of the approaches is that they cannot guarantee optimal or verifiable attack vector identification, stemming from their non-adherence to established grid and ADM constraints, unlike formal methods. Formal method-based analyzers synthesize verifiable attack vectors; however, existing efforts often fail to extract attack vectors from complicated/interdependent ML-based CPS dynamics in a feasible time. One of our notable research works, SHChecker~\cite{haque2021novel}, identifies optimal FDI attacks to provide wrong treatment to the connected healthcare systems bypassing different ML models~\cite{jafari2022optimal}. Another work, SHATTER, proficiently identifies attack vectors in ML-based smart home systems, focusing on time-series ADMs and considering multi-time slot attacks. However, such attack analysis frameworks can only identify attack vectors from independent controller dynamics. Jafari et al. identified FDI attacks by adopting a distinct approach and exploring the dynamics of complex power systems. Yet their research predominantly depends on a rules-based bad data detector (BDD). Throughout the write-up, we will comprehensively compare our proposed analyzer with this work, referred to as \textbf{TIFS'23}. While effective in some contexts, the strategy devised in TIFS'23 primarily identifies straightforward attack vectors subjected to be detected by the ML-based models. Hence, this highlights the need for more advanced and comprehensive attack analytics to effectively synthesize and verify complex attack vectors. The comparative analysis among the proposed and state-of-the-art approaches is highlighted in Table~\ref{tab:comparison-formal-adv_ml-rl-misguide}.

To address this research gap, we develop a novel attack analyzer naming \textbf{M}alicious-Activity \textbf{I}nvestigation for \textbf{S}mart \textbf{G}rid \textbf{U}tilizing \textbf{I}ntrusion \textbf{DE}tector (\framework). In this work, we designed and formally modeled a density-based spatial clustering of applications with Noise (DBSCAN)-based ADM that learns the benign time-series pattern of load measurements (i.e., load measurements perceived in the control center) to identify anomalous events. The proposed framework can identify a stealthy attack vector to optimally (i.e., in minimal time) inject malicious measurements into the load measurements perceived by the control center to trip protective relays (UF/OF), evading the ADM. \framework~is equipped to deal with the intricate dynamics of SG systems. The proposed analytics utilizes the Gurobi optimizer, overcoming the limitations of satisfiability modulo theories (SMT)-based solvers in handling complex mathematical equations and logical constraints. Hence, \framework~not only synthesizes sophisticated attack vectors but also ensures their verifiability. By covering a broader range of attack scenarios and intricacies in SG systems, \framework~significantly enhances the robustness of cybersecurity measures in these essential infrastructures. The load data for ADM training is obtained from state-of-the-art (SOTA) GEFCom2014 load forecasting dataset~\cite{hong2016probabilistic}. Our core contributions can be summarized as follows.
\begin{compactitem}
    \item We formally model the SG dynamics, including LFC, with an ML-based ADM, using first-order predicate logic by extracting constraints from the dependent component models to analyze the system.
    \item We develop an attack analytics to identify potential attack vectors in the communication packets between the generators and LFC by formally modeling FDI attacks with different attack attributes. The SOTA attack analytics (TIFS'23) investigated and identified FDI attacks based only on generator properties. In contrast, our proposed analyzer extends the analysis to identified attack vectors, considering diverse security properties, such as examining the ADM robustness, assessing attack space with different adversarial attributes, and analyzing system resiliency.
    \item We conduct experiments with our proposed analytics on SOTA real-world load pattern datasets of the ISO New England's IEEE 39-bus system to identify critical attack vectors and validate the identified attacks in a hardware-in-the-loop simulator (i.e., OPALRT). All implementation and evaluation results are reproducible with the source code on GitHub~\cite{misguide2024}.\footnote{Github repository: https://github.com/misguidetdsc/misguide}
\end{compactitem}

%%%%%%%%%%%%%%%%%%%%%%%%%%%%%%%
\begin{table}[!t]
\scriptsize
\centering
\caption{Summary of the Comparative Analysis of Formal Modeling, Adversarial ML, and RL Approaches for Attack Vector Identification from ML-based ADMs}
\label{tab:comparison-formal-adv_ml-rl-misguide}
\vspace{-6pt}
\begin{tabular}{|p{3.2cm}|p{0.9cm}|p{1.1cm}|p{0.4cm}|p{1.1cm}|}
\hline
\textbf{Criteria}                                                                              & \textbf{Formal Analysis}                    & \textbf{Adversarial ML}            & \textbf{RL} & \textbf{\framework}                     \\ \hline
Verifiable threat detection                                                           & $\checkmark$ & X                  & X & $\checkmark$\\ \hline
Solution identification guarantee                                                     & $\checkmark$ & X                  & X   & $\checkmark$               \\ \hline
Convergence                                                                           & $\checkmark$ & $\checkmark$ & X    & $\checkmark$              \\ \hline
Security-aware analysis & X                  & X                  & X & $\checkmark$\\ \hline
Feasible solution from large and complex time-series models & X                  & X                  & $\checkmark$ & $\checkmark$\\ \hline
\end{tabular}
\vspace{-9pt}
\end{table}

The rest of the paper is organized as follows: we provide an overview of the considered SG generators and LFC dynamics in Section~\ref{sec:sg-lfc}. We formally describe the problem domain and our considered attack model in Section~\ref{sec:problem-definition}. In the following section (i.e., Section~\ref{sec:technical-overview}), we present the technical overview of the proposed framework. We provide case studies to give insights about our proposed framework's working principle and capabilities in Section~\ref{sec:case-study}.
Then, we show the validation of the \framework~ with a real-time simulator. We evaluate \framework~using SOATA datasets in Section~\ref{sec:evaluation}. A comprehensive literature review is presented in Section~\ref{sec:related-work}. Section~\ref{sec:discussions} summarizes and discusses our research findings and limitations, and Section~\ref{sec:conclusion} concludes the write-up.
%%%%%%%%%%%%%%%%%%%%%%%%%%%%%
\vspace{-7pt}
%%%%%%%%%%%%%%%%%%%%%%%%%%%%%%%%%%%%%%%%%%%%%

%%%%%%%%%%%%%%%%%%%%%%%%%%%%%%%%%%%%%%%%%%%%%
\section{Smart Grid Load-Frequency Control System}
\label{sec:sg-lfc}

The LFC deals with critical mechanisms to balance electricity supply and demand within an SG. It ensures that the system frequency is kept within a specified range around a setpoint, which is essential for the stable operation of the grid. The control process involves adjusting the output of power generators in real time to respond to changes in load (electricity demand) and to maintain the frequency at its target value. Here, we discuss the dynamics of the LFC and the associated components, such as SG generator dynamics and protection relays. Moreover, we also incorporate the overview of the ADM considered for our proposed attack analytics.  

%%%%%%%%%%%%%%%%%%%%%%%%%%%%%%%%%%%
\begin{table}[!t]
\scriptsize
\centering
\caption{Modeling Notations}
\label{tab:modeling-notations}
\begin{tabular}{|l|l|p{4.4cm}|l|}
\hline
\textbf{\begin{tabular}[c]{@{}l@{}}SL.\\ No.\end{tabular}} & \textbf{Notation}                & \textbf{Description} & \textbf{\begin{tabular}[c]{@{}l@{}}Data\\ Type\end{tabular}} \\ \hline
%%%
1 & $\mathcal{B}$                    & Set of all buses in the system & Set \\ \hline
%%%
2 & $\mathcal{B}^{\mathit{PV}}$ & Set of generator (PV) buses in the system & Set \\ \hline
%%%
3 & $\mathcal{B}^{\mathit{GV}}$      & Set of PV buses in the system that have governors & Set \\ \hline
%%%
4 & $\mathcal{B}^{\mathit{PQ}}$ & Set of all load (PQ) buses in the system & Set \\ \hline
%%%
5 & $\mathcal{B}^{S}$ & Slack bus of the system & Integer \\ \hline
%%%
6 & $\mathcal{G}$ & Set of all generators & Set \\ \hline
7 & $\mathcal{T}$ & Set of all timeslots for primary controllers' response & Set \\ \hline
%%%
8 & $\mathcal{T}^C$ & Set of all timeslots for LFC's response ($\mathcal{T}^C$ $\subset$ $\mathcal{T}$) & Set \\ \hline
%%%
9  & $\mathcal{P}^{\mathit{L}}_{b,t}$, ($\mathcal{P}^{\mathit{L, C}}_{b,t}$) & Load active power of b-the bus at time, t (load measurement perceived in the LFC) & Real \\ \hline
%%%
10 & $\mathcal{P}^{\mathit{G}}_{b,t}$, ($\mathcal{P}^{\mathit{G, C}}_{b,t}$) & Electrical power (active) output of the generator connected to the b-th bus at the time, t, i.e., $b$ $\in$ $\mathcal{B}^{\mathit{PV}}$ (LFC estimated active power measurement) & Real \\ \hline
%%%
11 & $\mathcal{P}^{\mathit{M}}_{b,t}$, ($\mathcal{P}^{\mathit{M, C}}_{b,t}$) & Mechanical power output of the generator connected to the b-th bus at time, t i.e., $b$ $\in$ $\mathcal{B}^{\mathit{PV}}$ (LFC estimated mechanical power measurement) & Real \\ \hline
%%%
12 & $\mathcal{P}^{\mathit{R}}_{b,t}$ & Reference active power (p.u.) of the generator connected to the b-th bus at the time, t (i.e., $b$ $\in$ $\mathcal{B}^{\mathit{PV}}$) & Real \\ \hline
%%%
13 & $\delta_{b, t}$, ($\delta^C_{b, t}$) & Rotor angle of the generator connected to b-th bus at the time, t, i.e., $\delta_{b, t}$ = $\theta_{b, t}$ and $b$ $\in$ $\mathcal{B}^{\mathit{PV}}$ (LFC calculated phase angles) & Real \\ \hline
%%%
14 & $\omega_{b,t}$, ($\omega^C_{b,t}$)              & Angular frequency (p.u.) of the generator connected to the b-th bus at the time, t, i.e., $b$ $\in$ $\mathcal{B}^{\mathit{PV}}$ (LFC perceived angular frequency measurement) & Real \\ \hline
15 & $\omega^{R}$ & Nominal angular frequency of all buses & Real \\ \hline
%%%
16 & $\mathbb{R}_{b}$ & Droop co-efficient of b-th bus & Real \\ \hline
%%%
17 & $\Delta T$ & Discretization time-step size of primary controller's response time & Integer \\ \hline
%%%
%%%
18 & $\mathcal{T}_{b}$ & Time-constant of the generator connected to the b-th bus & Real \\ \hline
%%%
19 & $\mathbb{S}_{b1, b2}$                     & Imaginary part of line admittance (siemens) in between bus $b1$ and $b2$ & Real \\ \hline
%%%
%%%
20 & $\mathbb{H}_{b}$                & Inertia constant of the synchronous generator that is connected to the bus, b (i.e., $b$ $\in$ $\mathcal{B}^{\mathit{PV}}$) & Real \\ \hline
\end{tabular}
\vspace{-15pt}
\end{table}

%%%%%%%%%%%%%%%%%%%%%%%%%%%%%%%%%%%%%%%%
\vspace{-10pt}
\subsection{Grid and Controller Dynamics}

\begin{figure}[t]
\centering
\includegraphics[width=0.99\columnwidth]{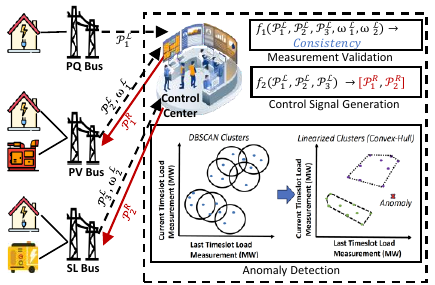}
\caption{Demonstration of an IEE-3 bus SG system.}
\label{fig:problem-overview}
\vspace{-15pt}
\end{figure}

To maintain grid frequency, generator governors continuously adjust power generation, ensuring that the frequencies of the generators remain synchronous. The LFC is a secondary controller that communicates with the SG generators and updates their reference setpoints periodically. A 3-bus SG system is illustrated in Figure~\ref{fig:problem-overview}, where the controller receives the load measurements and generates a control signal to set the generator setpoints, thus adjusting the system's frequency. The load (PQ) buses only host load components, while the PV and slack (SL) buses connect generators with or without loads. The working principle of primary and secondary frequency response is briefly described as follows.

\noindent \textbf{Primary Frequency Response:} SG's dynamic behavior is modeled using the differential equations of synchronous generators, governors, and linear power flow. When a power system changes, such as when there are fluctuations in the load, the power flow equation provides the updated rotor angle values, which aid the assessment of the frequency behavior of the system. SG dynamics are represented by non-linear differential-algebraic equations~\eqref{eq:swing-equation_delta}-\eqref{eq:dc_power_flow}. The modeling notations for power system components, dynamics, and FDI attacks are shown in Table~\ref{tab:modeling-notations}. The dynamic behavior of synchronous generators can be represented using the following swing equations~\eqref{eq:swing-equation_delta}-\eqref{eq:swing-equation_omega}. We consider the classical representation of synchronous generators in which the terminal voltage angles approximate the generator's rotor angles. The governor model TGOV1, the simplified representation of the steam governor, is
considered and can be represented by~\eqref{eq:governor-equation}. The linear power flow dynamics can be formalized as~\eqref{eq:dc_power_flow}. 
\vspace{-5pt}
%
%%%%%%%%%%%%%%%%%%%%%%
\begin{equation}
    \label{eq:swing-equation_delta}
    \begin{split}
    \forall_{b \in \mathcal{B}^{\mathit{PV}}} \frac{d\delta_b}{dt} = \omega_b - \omega^R = \Delta \omega_b
    \end{split}
\end{equation}
%%%%%%%%%%%%%%%%%%%%%%
\vspace{-10pt}
%
%%%%%%%%%%%%%%%%%%%%%%
\begin{equation}
    \label{eq:swing-equation_omega}
    \begin{split}
    \forall_{b \in \mathcal{B}^{\mathit{PV}}}\frac{d\omega_b}{dt} = \frac{1}{2\mathbb{H}_b} ( \mathcal{P}^M_b - \mathcal{P}^G_b - \mathbb{K}^D_b \Delta \omega_b )
    \end{split}
\end{equation}
%%%%%%%%%%%%%%%%%%%%%%
\vspace{-10pt}
%
%%%%%%%%%%%%%%%%%%%%%%
\begin{equation}
    \label{eq:governor-equation}
    \begin{split}
    \forall_{b \in \mathcal{B}^{\mathit{PV}}}\mathcal{P}^M_b = \frac{1}{\mathbb{T}_b}\int \left( \frac{ \mathcal{P}^R_b - \Delta\omega_b}{\mathbb{R}_b} \right)  - \mathcal{P}^M_b ) dt
    \end{split}
\end{equation}
%%%%%%%%%%%%%%%%%%%%%%
\vspace{-10pt}
%
%%%%%%%%%%%%%%%%%%%%%%
\begin{equation}
    \label{eq:dc_power_flow}
    \begin{split}
    \forall_{\mathit{b1} \in \mathcal{B}} \mathcal{P}^G_\mathit{b1} - \mathcal{P}^L_\mathit{b1} = \sum_{b2 \in \mathcal{B}} 
    \mathbb{S}_\mathit{b1,b2}
    (\mathcal{\delta}_\mathit{b1}-\mathcal{\delta}_\mathit{b2})
    \end{split}
\end{equation}
%%%%%%%%%%%%%%%%%%%%%%
%
%
\noindent \textbf{Secondary Frequency Control:} The secondary controller/LFC receives the load measurements of all the buses and frequency measurements of the generators with governors. The frequency measurement calculates the rotor angle of the generators shown in~\eqref{eq:swing-equation_delta_lfc}. The control center runs the state estimations (as in~\eqref{eq:dc_power_flow_lfc}) based on estimated generator rotor angles and received load measurements (e.g., $\mathcal{P}^{\mathit{L, C}}$ as shown in Figure~\ref{fig:lfc}) and determine the generators reference setpoints (dispatch signal) $\mathcal{P}^{\mathit{R, C}}$ as~\eqref{eq:dispatch-signal-lfc}.
\vspace{-5pt}
%%%%%%%%%%%%%%%%%%%%%%
\begin{equation}
    \label{eq:swing-equation_delta_lfc}
    \begin{split}
    \forall_{b \in \mathcal{B}^{\mathit{PV}}} \frac{d\delta_b}{dt} = \omega^C_b - \omega^R
    \end{split}
\end{equation}
%%%%%%%%%%%%%%%%%%%%%%
\vspace{-10pt}
%
%%%%%%%%%%%%%%%%%%%%%%
\begin{equation}
    \label{eq:dc_power_flow_lfc}
    \begin{split}
    \forall_{\mathit{b1} \in \mathcal{B}} \mathcal{P}^{\mathit{G, C}}_\mathit{b1} - \mathcal{P}^{\mathit{L, C}}_\mathit{b1} = \sum_{b2 \in \mathcal{B}} 
    \mathbb{S}_\mathit{b1,b2}
    (\mathcal{\delta}^C_\mathit{b1}-\mathcal{\delta}^C_\mathit{b2})
    \end{split}
\end{equation}
%%%%%%%%%%%%%%%%%%%%%%
\vspace{-10pt}
%%%%%%%%%%%%%%%%%%%%%%
\begin{equation}
\label{eq:dispatch-signal-lfc}
    \begin{split}
    \forall_{b \in \mathcal{B}^\mathit{PV} } \mathcal{P}^\mathit{R}_b = \mathbb{R}_b \mathcal{P}^\mathit{G, C}_b
    \end{split}
\end{equation}
%%%%%%%%%%%%%%%%%%%%%%
%
%
To illustrate the general characteristics of the controller dynamics, we omit the time index in these equations. The discretized equations, including time indices, are presented in Section~\ref{subsec:grid-dynamics-constraints}.

\vspace{-15pt}
%%%%%%%%%%%%%%%%%%%%%%%%%%%%%%%%%%
\subsection{Protection Relays}
Smart grid UF and OF protection relays are crucial components in the electrical power system, ensuring stability and reliability. They are triggered when the power system's frequency deviates from the nominal frequency beyond predetermined thresholds. OF relays activate when the frequency exceeds a set point, indicating excess generation or reduced load. In contrast, UF relays respond to a drop below a set point, signaling insufficient generation or excessive load.

%%%%%%%%%%%%%%%%%%%%%%
%%%%%%%%%%%%%%%%%%%%%%
%%%%%%%%%%%%%%%%%%%%%%
%%%%%%%%%%%%%%%%%%%%%%

%%%%%%%%%%%%%%%%%%%%%%%%%%%%%%%%%%
\vspace{-10pt}
\subsection{Measurement Validation and Anomaly Detection}
TIFS'23 demonstrated that FDI attacks on the SG communication systems could inadvertently actuate protection relays, resulting in the disconnection of a substantial load or a large generator from the system~\cite{jafari2023optimal}. The identified attack vectors successfully evaded BDD rules. The BDD rules were designed to monitor deviations in load measurements across successive control cycles, with a significant deviation triggering an alarm and enabling system administrators to implement preventative measures. However, the introduction of stealthy attack vectors and the prevalence of advanced persistent threats demonstrates the potential to activate UF or OF relays without setting off any alarms. In our work, we envision a more resilient system that learns the patterns of allowable deviations from historical load data, not solely depending on domain knowledge. We utilize clustering-based ML models to train the ADM, which learns the dynamic threshold between consecutive load measurements. Our considered resiliency in the SG LFC has two core functional components - measurement validation and anomaly detection.

\noindent \textbf{Measurement Validation:} We already identified the relationship between the load measurements and different generator parameters as shown in~\eqref{eq:swing-equation_delta_lfc}-\eqref{eq:dispatch-signal-lfc}. Intuitively, if the load measurements in consecutive control cycles do not vary significantly, but the frequency measurements of the generators fluctuate drastically, this is considered an invalid set of measurements.

\noindent \textbf{Anomaly Detection:} The functionality of the ADM is showcased in Figure~\ref{fig:problem-overview}. We consider different clustering-based ML models for learning all load measurements, i.e., each ML model learns the load patterns for a particular bus. The consistency of the load measurements is checked for all ML models, i.e., all load measurements perceived in the LFC should be inside at least one of the clusters of all the ML models. Otherwise, the ADM will detect the phenomena as an anomaly/attack event. 

At the control center, the ADM learns the patterns of benign load measurements, represented as clusters trained on the dynamic relationship between load measurements from consecutive time slots. For instance, if the previous load measurement from bus 9 was 5 MW, the current measurement is expected to fall within a range of 4.9-5.1 MW. Each cluster produced by the Density-Based Spatial Clustering of Applications with Noise (DBSCAN) algorithm forms multiple regions later linearized into a set of convex hulls. Training points are used instead of non-linear boundaries to generate these convex hulls. This approach ensures that the convex hull does not extend beyond the actual ADM boundary, minimizing the risk of creating attack vectors that could be detectable by the ADM.
%%%%%%%%%%%%%%%%%%%%%%%%%%%%%
%%%%%%%%%%%%%%%%%%%%%%%%%%%%%%%%%%%%%%%%%%%%%
%%%%%%%%%%%%%%%%%%%%%%%%%%%%%%%%%%%%%%%%%%%%%
\section{Problem Definition and Attack Model}
\label{sec:problem-definition}

This section provides a formal definition of the LFC dynamics and summary of the attack model. 

\vspace{-10pt}
%%%%%%%%%%%%%%%%%%%%%%%%%%%%%%%
\subsection{Problem Definition}
An smart power grid $\mathcal{P}$ is considered aiming to satisfy the demands/loads ($\mathcal{L}$) of component bus $\mathcal{B}$ through the synchronous generators $\mathcal{G}$ connected with some of those buses. The impact of overproduction or underproduction can be observed by the generators' frequency measurements (i.e., $\mathcal{F}$ = $\frac{\omega}{2\pi}$). The governor of the generators adjusts generation based on the deviation of expected and current angular frequency and the reference setpoint of the generators. The LFC regulates the reference setpoints of the generators. The control center periodically receives load measurements $\mathcal{P}^L$ of $\mathcal{B}$ and dispatches the reference setpoint $\mathcal{P}^R$ of the generators through a controller. Since power grids require fast communication of measurements, the measurements cannot be ciphered with high-complexity encryption algorithms, making them susceptible to attacks. Hence, the load measurements perceived in the LFC can be misleading since the measurements can be altered in the communication line of the generators to the LFC.

% %%%%%%%%%%%%%%%%%%%%%%%%%%%%%%%%%%
% \begin{definition}[Attack Vector]
% \label{def:attack-vector}
% The malicious measurements added to the actual measurements 
% \end{definition}

To deal with the problem, an ML-based ADM $\mathbb{E}$ checks the load measurement consistencies of the consecutive LFC cycle load measurements learning through years of load measurements of the system. The ADM will detect an FDI attack intended to alter the load measurements arbitrarily or based on some rules. A stealthy attack can be launched if the attacker intelligently changes the load measurements within the range of the ADM rules. The main challenge of identifying attack vectors that can attain the attack goal of triggering UF or OF relays evading the dynamic threshold is provided by $\mathbb{E}$.

\vspace{-10pt}
%%%%%%%%%%%%%%%%%%%%%%%%%%%%%%
\subsection{Attack Model}
The attack model generates parameterized attack procedures and functions that target a specific CPS, in our case, an SG. This section provides a summarized version of the attack model, detailed and formally analyzed in Section~\ref{sec:technical-overview}.

%%%%%%%%%%%%%%%%%%%%%%%%%%%%%%%%%%
\subsubsection{Attack Strategies}
Completing an FDI attack in the presented context requires acquiring the ADM parameters and real-time measurements. Once the measurements are received, they must be altered following the ADM. %The principal task of the framework can be 

\noindent \textbf{Measurement Acquisition:} The first phase of carrying out an FDI attack is to get access to the measurements passively. The required measurements to be sniffed include the load measurements of different buses and frequency measurements of the generators. Measurement acquisition can be performed through physical sensing. Low-cost sniffers enable reconnaissance attacks that continuously monitor the measurements~\cite{zhu2018tu}. Eavesdropping communication packets is another method where attackers can position as a man-in-the-middle (MITM) and sniff communication packets using packet capture and analysis tools. 

\noindent \textbf{Measurement Alteration:} The second or the active attack phase is to tamper sensor measurements, performed mainly by packet crafting using techniques like ARP poisoning and IP/MAC addressing spoofing~\cite{ding2018safety}.

%%%%%%%%%%%%%%%%%%%%%%%%%%%%%%%%%%
\subsubsection{Attack Assumptions}
The proposed framework considers a set of assumptions.
\begin{compactenum}[(a)]
    \item \textit{Assumption I:} The attacker has complete knowledge of the LFC properties, control algorithm, and ADM. Moreover, the parameters considered in the generators are known to the attacker.
    \item \textit{Assumption II:} The attackers maintain complete and limited access to all sensor measurements, such as generators' and buses' load and frequency measurements. However, the dispatched control signals sent to the generators are considered to be protected against FDI attacks.
    \item \textit{Assumption III:} The benign load measurements considered in the attack period are constant.
    \item \textit{Assumption IV:} The attack model considers that the attacker cannot modify the measurements indefinitely; thus, the duration of measurement alteration is constrained. Our attack model considers accessibility to sensor measurement to imply that an attacker can read and modify a sensor measurement.
    \item \textit{Assumption V:} The attacker's access to the load measurements suggests they also have access to the angular frequency measurements of the corresponding bus.
\end{compactenum}

%%%%%%%%%%%%%%%%%%%%%%%%%%%%%%%%%%
\subsubsection{Adversarial Attributes}
The attacker cannot arbitrarily inject measurements. Two adversarial attributes- the attacker's accessibility and capability determine the attack intensity.
\textit{\ul{Attacker's accessibility}} pertains to the opportunity or how an attacker can reach a system or asset. It determines what the attacker has potential access to and could be influenced by several factors, such as network configurations, exposed services, perimeter defenses, and other security controls. The attacker's accessibility can be represented using a boolean vector $\mathbb{A}$. If an element in the accessibility matrix is indicated false, the corresponding bus load measurement cannot be attacked/altered. \textit{\ul{Attacker's capability}} pertains to an attacker's skills, resources, tools, and knowledge. It determines what the attacker can potentially do. As mentioned in the previous section, measurement alteration capability is limited for launching a stealthy attack since the altered measurements need to be consistent with the ADM.

%%%%%%%%%%%%%%%%%%%%%%%%%%%%%%%%%%
\subsubsection{Attack Goal}
In this work, we consider the attacker's goal is to trigger an OF or UF relay of at least one generator or load. Activation of a relay often has a cascading impact on the system, especially for the OF relays. However, our scope is limited to assess the attack impact by triggering relays.

%%%%%%%%%%%%%%%%%%%%%%%%%%%%%
%%%%%%%%%%%%%%%%%%%%%%%%%%%%%%%%
%%%%%%%%%%%%%%%%%%%%%%%%%%%%%%%%
\section{Detailed Technical Overview of \framework~Framework}
\label{sec:technical-overview}

This section provides a technical outline of the functionality of the proposed analytics.
%%%%%%%%%%%%%%%%%%%%%%%%%%%%%%%%

\vspace{-12pt}
\subsection{Formal Modeling of Grid Dynamics Constraints}
\label{subsec:grid-dynamics-constraints}

The dynamic behavior of power systems is modeled by observing the changes in synchronous generators’ rotor angles and assessing power systems’ frequency stability. The linear power flow equations are used, which gives us the new values of rotor angles for any changes in power systems, such as generators’ active power output or the loads. However, the continuous format is not implementable for the SMT-based solver. We discretize these equations using the Backward Euler method to overcome this issue. The discretized version of swing equations is depicted in~\eqref{eq:delta_update}~and~~\eqref{eq:omega_update}. The discretized version of the governer equation is shown in~\eqref{eq:mechanical_power_update}. 

The control center calculated linear power flow dynamics can be formalized in a discretized version as~\eqref{eq:dc_power_flow_update}. Based on the power flow equation, the generators’ reference setpoints are adjusted based on~\eqref{eq:reference_setpoint_update}.  
%
%%%%%%%%%%%%%%%%%%%%%%
\begin{equation}
    \label{eq:delta_update}
    \begin{split}
        \forall_{b \in \mathcal{B}^\mathit{PV}, t \in \mathcal{T}} 
 \delta_\mathit{b,t+1} = \mathcal{\delta}_\mathit{b,t} + \Delta\mathcal{T} (\omega_\mathit{b,t+1} - \omega^R)
    \end{split}
\end{equation}
%%%%%%%%%%%%%%%%%%%%%%
\vspace{-10pt}
%
%%%%%%%%%%%%%%%%%%%%%%
\begin{equation}
    \label{eq:omega_update}
    \begin{split}
        \forall_{b \in \mathcal{B}^{PV}, t \in \mathcal{T}} \omega_\mathit{b,t+1} = \omega_\mathit{b,t} + \frac{\Delta T}{2 \mathbb{H}_{b}} (\mathcal{P}^M_{b,t+1} - \mathcal{P}^G_{b,t+1} \\ \mathbb{K}^D_b (\omega_\mathit{b,t+1} - \omega^R))
    \end{split}
\end{equation}
%%%%%%%%%%%%%%%%%%%%%%
\vspace{-8pt}
%
%%%%%%%%%%%%%%%%%%%%%%
\begin{equation}
    \label{eq:mechanical_power_update}
    \begin{split}
        \forall_{b \in \mathcal{B}^{PV}, t \in \mathcal{T}} 
        \mathcal{P}^M_\mathit{b,t+1} =
         \mathcal{P}^M_\mathit{b,t} + \frac{\Delta T}{\mathbb{T}_{b} \mathbb{R}_b} \times (\mathcal{P}^R_\mathit{b,t+1} - (\omega_\mathit{b,t+1} \\ - \omega^R)) - \mathbb{R}_{b} \mathcal{P}^M_\mathit{b,t+1}
    \end{split}
\end{equation}
%%%%%%%%%%%%%%%%%%%%%%
\vspace{-8pt}
%
%%%%%%%%%%%%%%%%%%%%%%
\begin{equation}
    \label{eq:dc_power_flow_update}
    \begin{split}
    \forall_{b1 \in \mathcal{B}, t \in \mathcal{T}} \mathcal{P}^G_\mathit{b_1,t} - \mathcal{P}^L_\mathit{b_1,t} = \sum_{b2 \in \mathcal{B}} 
    \mathbb{S}_{b1,b2}
    (\delta_\mathit{b1,t} - \delta_\mathit{b2,t})
    \end{split}
\end{equation}
\vspace{-10pt}
%%%%%%%%%%%%%%%%%%%%%%
\begin{equation}
    \label{eq:reference_setpoint_update}
    \begin{split}
        \forall_{b \in \mathcal{B}^{PV}, t \in \mathcal{T}^S} \mathcal{P}^R_\mathit{b,t} = \mathbb{R}_\mathit{b,t} \mathcal{P}^{G, C}_\mathit{b,t}
    \end{split}
\end{equation}
%%%%%%%%%%%%%%%%%%%%%%
%
%
As mentioned, a DBSCAN-based ADM learns from the benign load pattern to identify anomalous data in real time. The key idea is that DBSCAN forms clusters learning the relationship between previous and current load measurements. Although multiple ML models are needed to capture the load measurement relationships for each bus, we discuss only a single ML model for a generic bus here for the sake of brevity. We denote the convex hull of the clusters using the formal notation $\mathcal{C}$, where $\mathbb{N}^C$ = $|\mathcal{C}|$ represents the number of clusters. For formal modeling purposes, we linearize the clusters using a convex hull. The vertices in a cluster hull are denoted by a notation $\mathcal{X}_c$, while a particular vertice point p is expressed as  $\mathcal{X}_\mathit{c, p}$. Number of vertices in cluster is represented as $\mathbb{N}^P_c$ = $|\mathcal{X}_c|$. Note that the cluster vertices are arranged in an anticlockwise order. Suppose the clusters are formed using $l$ previous load measurements. Hence, including the current load measurement, the dimension of cluster points becomes $l+1$. The clusters’ convex hull can be expressed as a set of hyper-planes that connects each two consecutive anticlockwise considered component vertices. The set of hyperplanes can be denoted as $\mathcal{H}$. In contrast, a particular hyperplane q of a cluster c is expressed as $\mathcal{H}_\mathit{c, q}$ (i.e., $\mathcal{H}_\mathit{c, q}$), which is represented by coefficients (i.e., $\alpha_\mathit{c, h}$). The number of hyperplanes in a cluster depends on the degree and number of vertices. The number of hyperplanes in cluster is represented as $\mathbb{N}^H_c$ = $|\mathcal{H}_c| = f(\mathcal{X}_c)$. A point ($\mathcal{P}^\mathit{L,C}_\mathit{t-l}, ..., \mathcal{P}^\mathit{L,C}_\mathit{t}$) is considered bad data if it falls into clusters identified by the clustering algorithm. The condition for the current load measurements to be consistent with the previous ones is demonstrated in~\eqref{eq:point-inside-cluster}.
%
%%%%%%%%%%%%%%%%%%%%%%
\begin{equation}
    \label{eq:point-inside-cluster}
    \begin{split}
        \exists!_{c \in \mathcal{C}}\forall_{t \in \mathcal{T}^S, h \in \mathcal{H}_c} \sum_{i = 0}^l \alpha_\mathit{c,h,i} \mathcal{P}^{\mathit{L, C}}_\mathit{t-l-i} \leq 0
    \end{split}
\end{equation}
%%%%%%%%%%%%%%%%%%%%%%%%%%%%%%%%%%%%%%%%%%%%%%%%%%%%%%

%%%%%%%%%%%%%%%%%% Case Study 1 %%%%%%%%%%%%%%%%%%%%
%%%%%%%%%%%%%%%%%%%%%%%%%%%%%%%%%%%%%%%%%%%%%%
\begin{figure*}[!t]
    \begin{center}
        \subfigure[]
        {
        \label{case-study-1-load}
            \includegraphics[width=0.48\columnwidth]{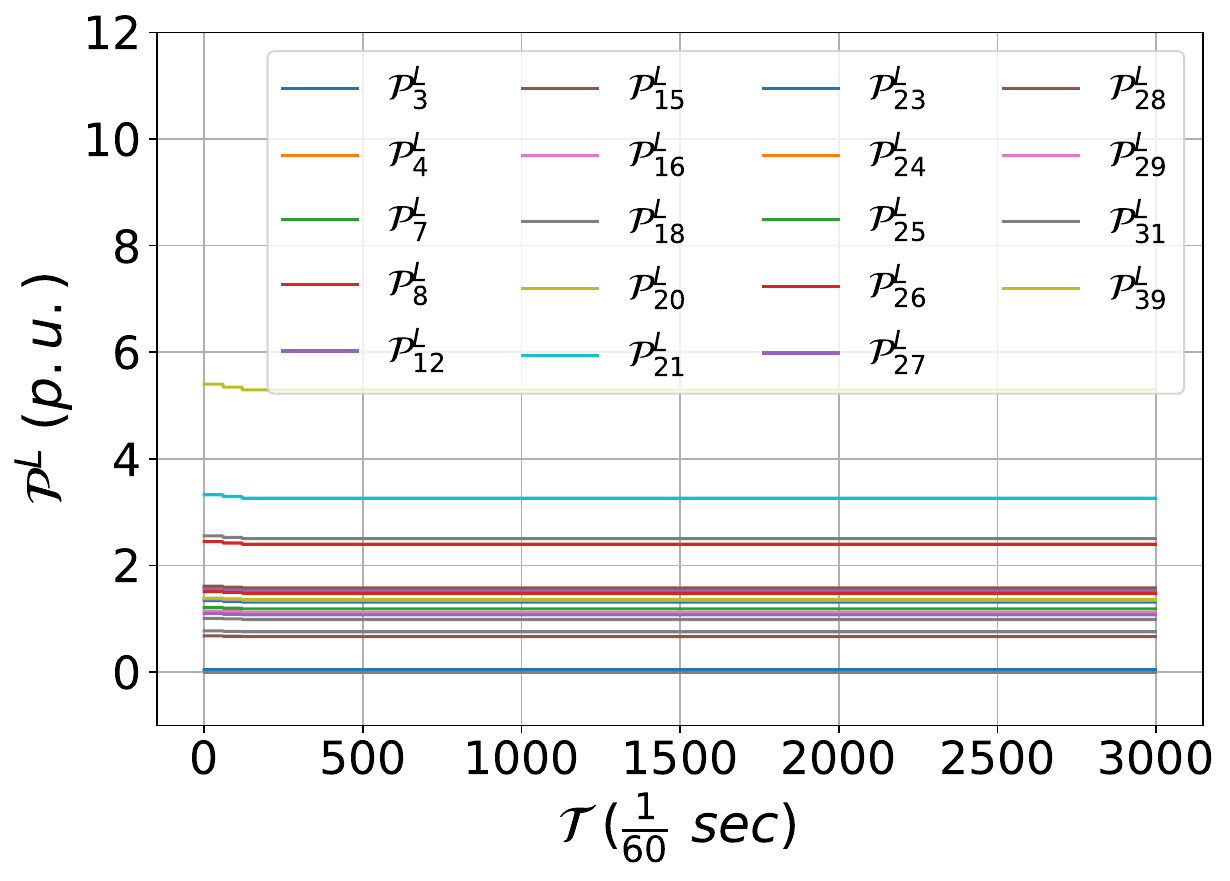}
        }
        \hspace{-10pt}
        \subfigure[]
        {
        \label{case-study-1-pg}
            \includegraphics[width=0.47\columnwidth]{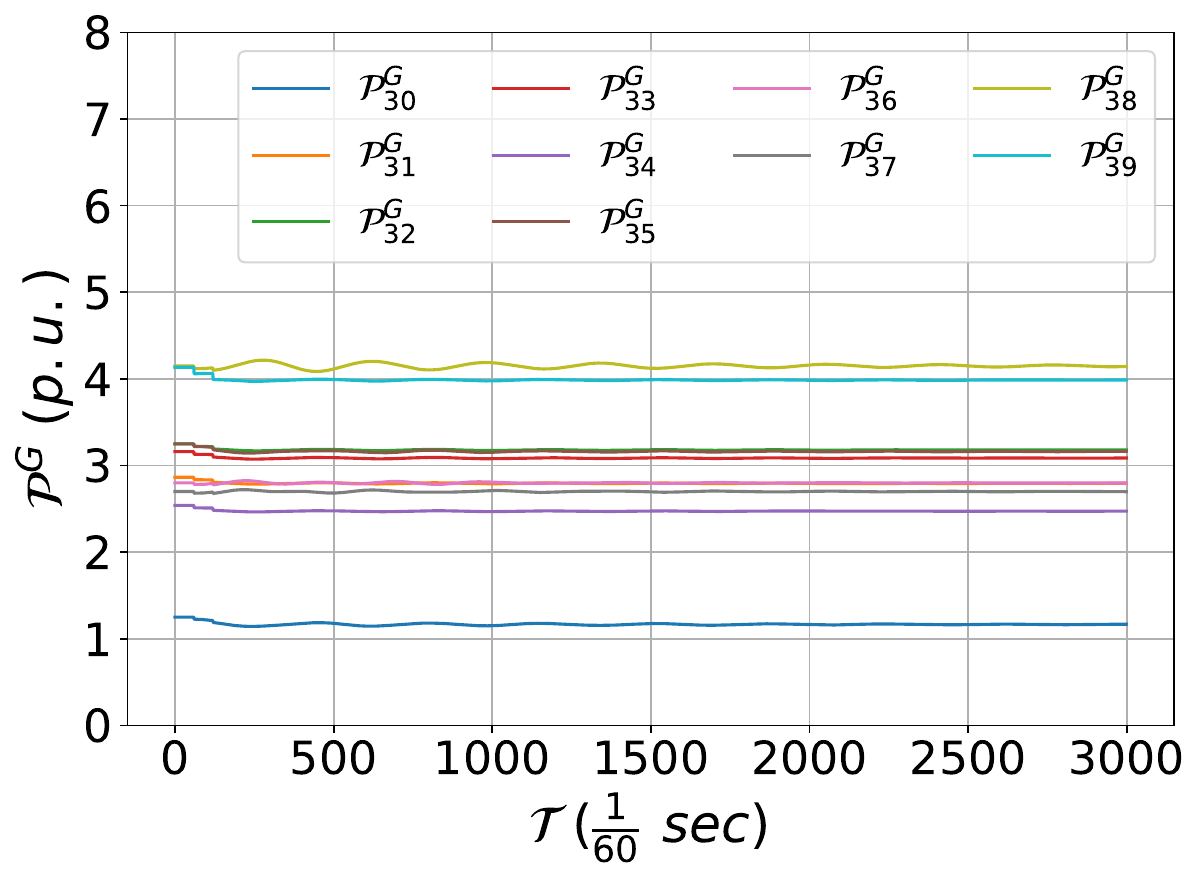}
        }
        \hspace{-10pt}
        \subfigure[]
         {
        \label{case-study-1-pr}
            \includegraphics[width=0.51\columnwidth]{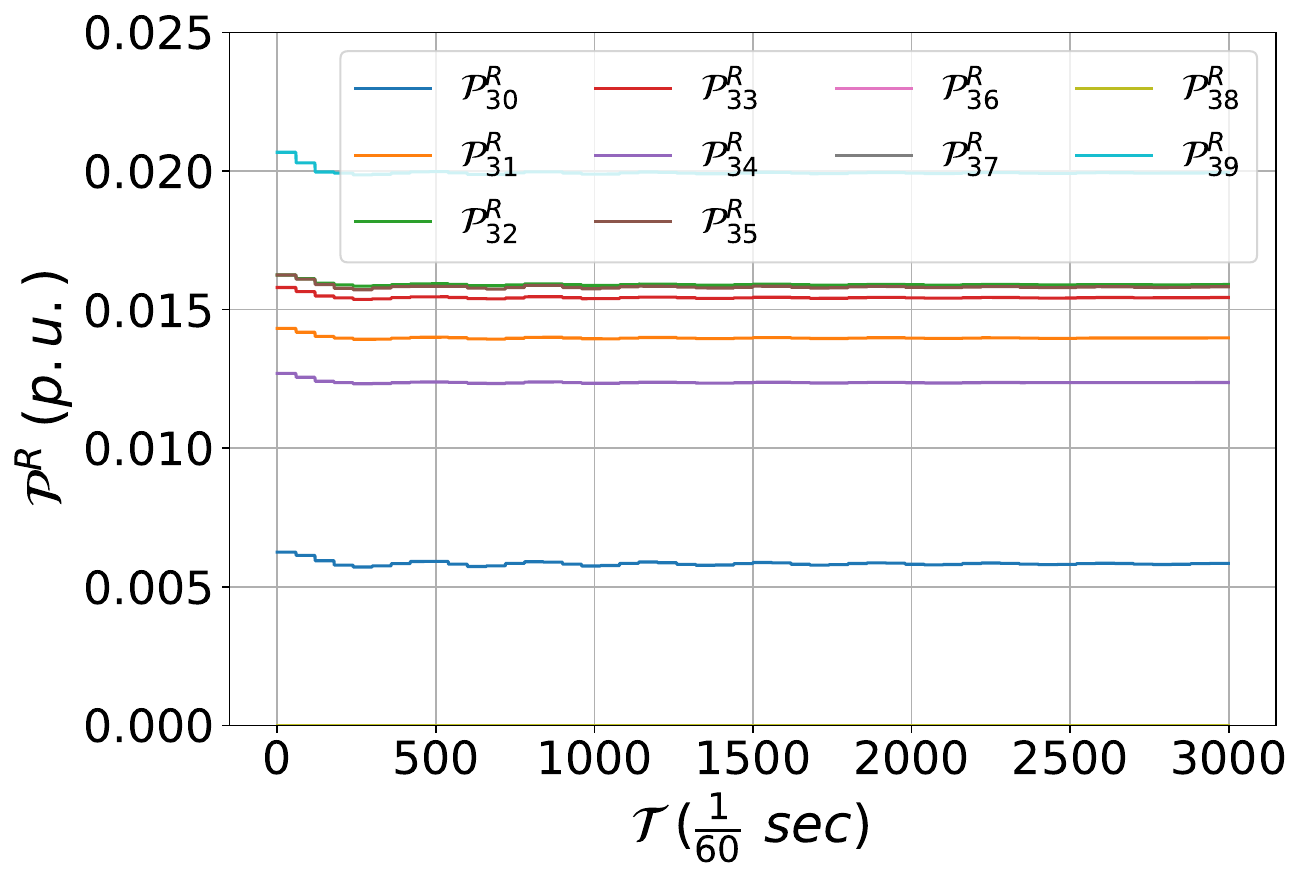}
        }
        \hspace{-10pt}
        \subfigure[]
         {
        \label{case-study-1-f}
            \includegraphics[width=0.5\columnwidth]{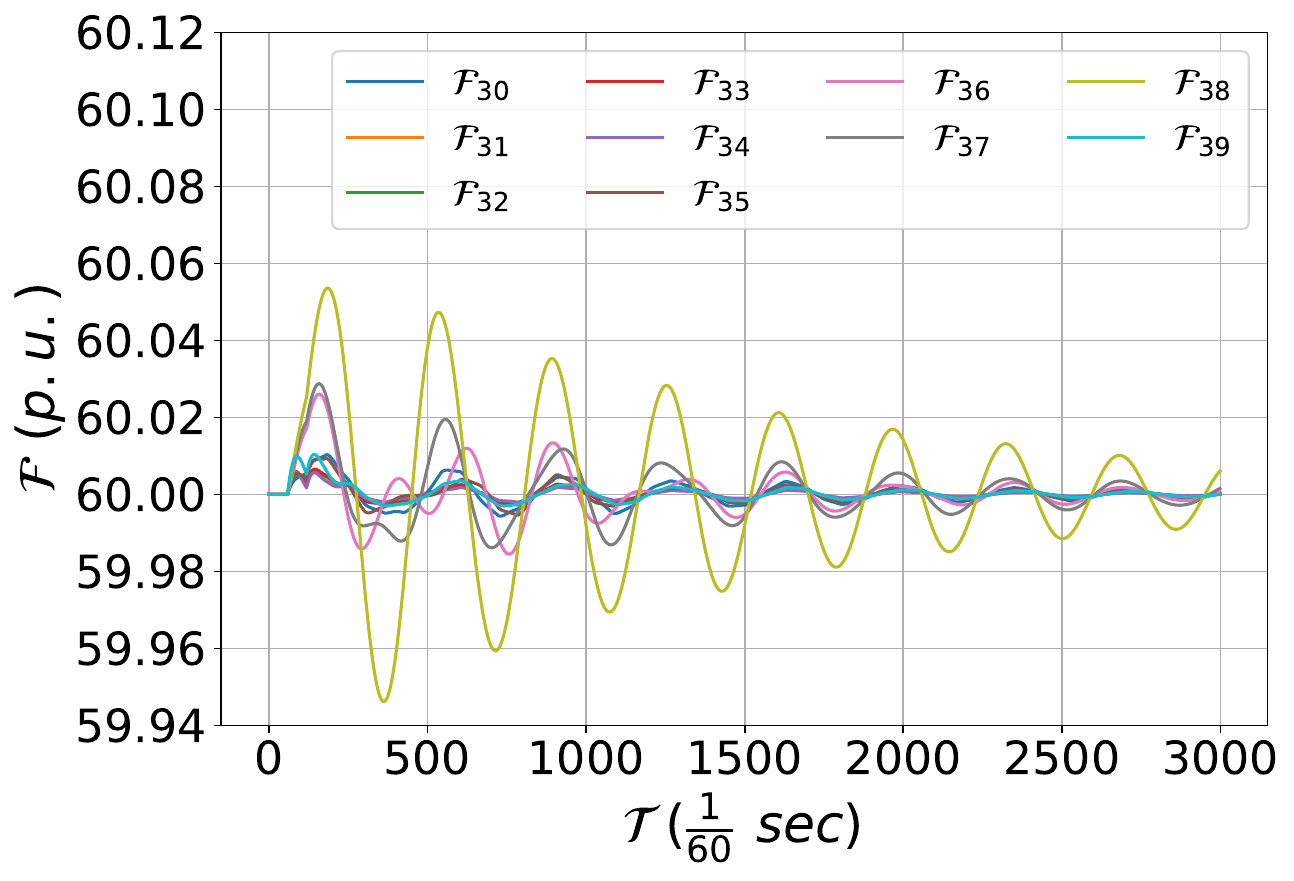}
        }
        %\hspace{-10pt}
    \end{center}
    \vspace{-15pt}
    \caption{\small Demonstrating the benign (a) load measurements (p.u.) of the buses and (b) generated active power (p.u.), (c) reference setpoint (p.u.), and (d) frequency (Hz) of different SG generators.}
    \label{fig:case-study-1}
    %\vspace{-6pt}
\end{figure*}

\subsection{Formal Modeling of FDI Attack Constraints}
Here, we talk about the attack technique, goal, and adversarial attributes to launch an optimal FDI attack to inject malicious measurements into the LFC perceived total load measurement to trip protective relays in minimal time. The optimal FDI attack compromises the load measurements, taking the power grid’s dynamics and the control center’s responses into account.

%%%%%%%%%%%%%%%%%%
\subsubsection{Attack Technique}
Assuming $\tilde{\mathcal{P}^{L,C}_{b,t}}$ is the attack vector injected into the load measurement, and $\bar{\mathcal{P}^{L,C}_{b,t}}$ are the attacked measurements, the FDI attack can be formally modeled as~\eqref{eq:load-attack-technique}.
\begin{equation}
    \label{eq:load-attack-technique}
    \begin{split}
    \forall_{b \in \mathcal{B}, t \in \mathcal{T}^S} \bar{\mathcal{P}}^{L,C}_{b,t} = \mathcal{P}^{L,C}_{b,t} + \tilde{ \mathcal{P}}^{L,C}_{b,t}
    \end{split}
\end{equation}
%

%%%%%%%%%%%%%%%%%
\subsubsection{Attack Goal}
The attack goal is to trigger FRO by injecting false measurements into the bus load measurements. The attack is deemed successful if at least one protection relay is maliciously triggered as shown in~\eqref{eq:load-attack-goal}.
\begin{equation}
\label{eq:load-attack-goal}
\exists_{t \in \mathcal{T}^S, b \in \mathcal{B}^\mathit{PV}}  (\omega_\mathit{b, t} \leq \mathbb{T}^\mathit{UF}) \lor (\omega_\mathit{b, t} \geq \mathbb{T}^\mathit{OF})
\end{equation}
%%%%%%%%%%%%%%%%%%%%%%%%%%%%%%%%%%%%%%%%%%%%%%%%%%%%%%%%%

%%%%%%%%%%%%%%%%%%%%%%%%%%%%%%%%%%%%%%%%%%%%%%%%%%%%%%%%%
\subsubsection{Adversarial Attributes}
The attacker cannot arbitrarily inject measurements. Two adversarial attributes- the attacker’s accessibility and capability determine the attack intensity.
\textbf{Attacker’s Accessibility} pertains to the opportunity or how an attacker can reach a system or asset. It determines what the attacker has potential access to and could be influenced by several factors, such as network configurations, exposed services, perimeter defenses, and other security controls. The attacker’s accessibility can be represented using a boolean vector $\mathbb{A} \in \mathbb{R}^{|\mathcal{B}|}$. If an element in the accessibility vector is indicated false, the corresponding bus load measurement cannot be attacked/altered.
\begin{equation}
    \label{eq:attacker-accessibility-load-measuements}
    \begin{split}
    \forall_{b \in \mathcal{B}, t \in \mathcal{T}^S} \mathbb{A}_b = 0 \rightarrow \tilde{ \mathcal{P}}^{L,C}_{b,t} = 0
    \end{split}
\end{equation}
\textbf{Attacker’s Capability} pertains to an attacker's skills, resources, tools, and knowledge. It determines what the attacker can potentially do. As mentioned in the previous section, measurement alteration capability is limited for launching a stealthy attack since the altered measurements need to be consistent with~\eqref{eq:point-inside-cluster}.

\subsection{Handling Logical Constraints for Gurobi Optimizer}
Gurobi is an optimization solver designed to address complex mathematical programming problems. However, Gurobi does not directly support logical constraints on variables. Three logical operations incorporate logical constraints into Gurobi: implication, logical and, and logical or. Although Gurobi cannot directly handle the implication operation, it can manage it indirectly through other means.

\begin{definition}[Indicator Constraint]
An indicator constraint, written as $i = v \rightarrow Ax + b < c$, indicates that if the binary indicator variable $i$ is set to $v$, then the linear constraint $Ax + b < c$ must be satisfied. Conversely, the linear constraint may be violated if $i$ is set to $1 - v$.
\end{definition}

The indicator constraint resembles the implication constraint but requires Boolean variables on the left-hand side. However, to handle conditional statements such as ``if $a > \mathbb{A}_1$ then $c = \mathbb{A}_2$ else $c = \mathbb{A}_3$," which can be expressed as $a > \mathbb{A}_1 \rightarrow c = \mathbb{A}_2$ and $a \leq \mathbb{A}_1 \rightarrow c = \mathbb{A}_3$, implication constraints are inadequate if $a$ is not Boolean. To model such logic in Gurobi, an auxiliary binary variable $b$ and indicator constraints are introduced.

In Gurobi, to model $a > \mathbb{A}_1$, we can use~\cite{gurobilogicalconst2024}:
\begin{equation*}
    a > \mathbb{A}_1 + \epsilon - M(1 - b)
\end{equation*}
\begin{equation*}
    a \leq \mathbb{A}_1 + M b
\end{equation*}
$\epsilon$ is a small positive tolerance to simulate the $a >$ constraint and $M$ is a sufficiently large number to avoid numerical issues. The binary variable $b$ then allows us to formulate the indicator constraints $b = 1 \rightarrow c = \mathbb{A}_2$ and $b = 0 \rightarrow c = \mathbb{A}_3$.

For handling logical operations such as logical and or logical or, Gurobi provides the functions $addGenConstrAnd()$ and $addGenConstrOr()$. These functions work with Boolean parameters only. Therefore, for non-Boolean linear expressions, we use intermediate Boolean variables to encode these operations similarly to how we handle implication operations.

%%%%%%%%%%%%%%%%%%%%%%%%%%%%%
%%%%%%%%%%%%%%%%%%%%%%%%%%%%%%%%%%%%%%%%%%%%
%%%%%%%%%%%%%%%%%%%%%%%%%%%%%%%%%%%%%%%%%%%%
\section{Case Studies}
\label{sec:case-study}

%%%%%%%%%%%%%%%%%% Case Study 2 %%%%%%%%%%%%%%%%%%%%
%%%%%%%%%%%%%%%%%%%%%%%%%%%%%%%%%%%%%%%%%%%%%%
\begin{figure*}[!t]
    \begin{center}
        \subfigure[]
        {
        \label{case-study-2-attacked-load}
            \includegraphics[width=0.45\columnwidth]{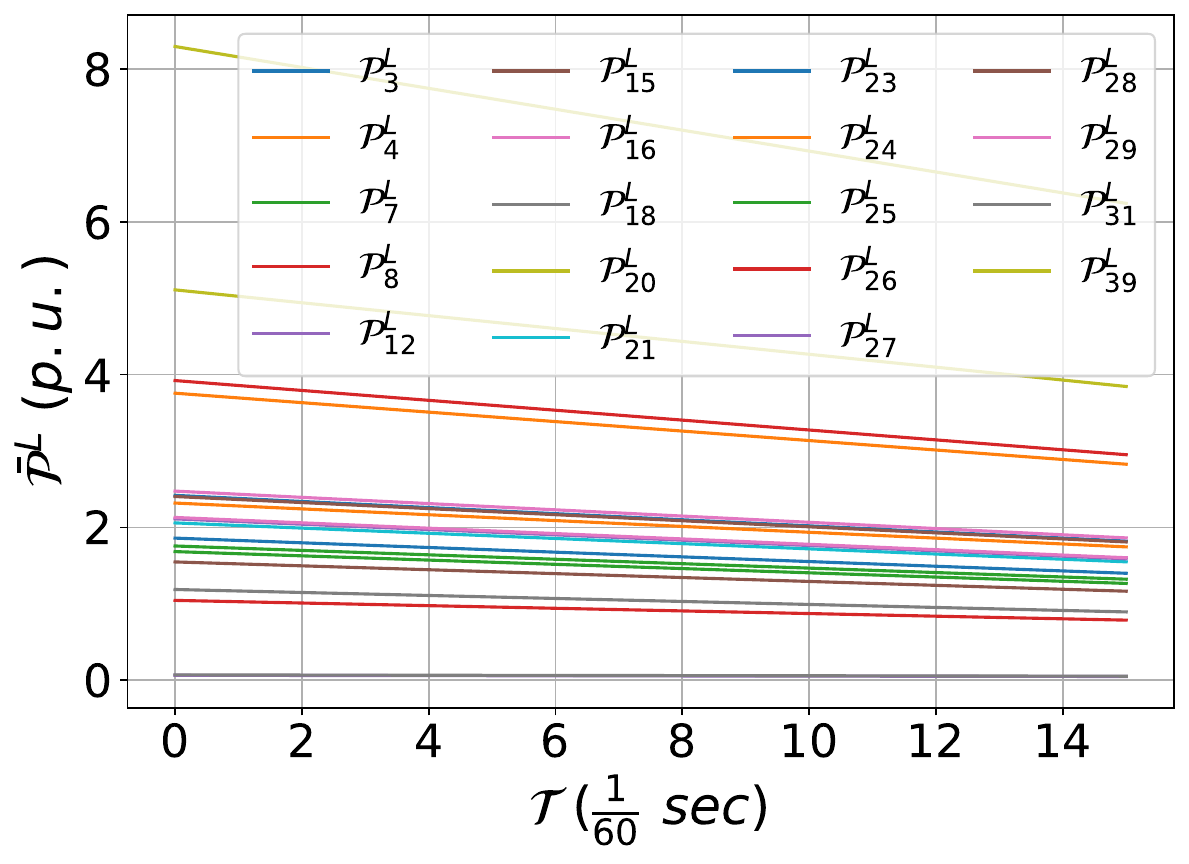}
        }
        \hspace{-10pt}
        \subfigure[]
        {
        \label{case-study-2-pg}
            \includegraphics[width=0.49\columnwidth]{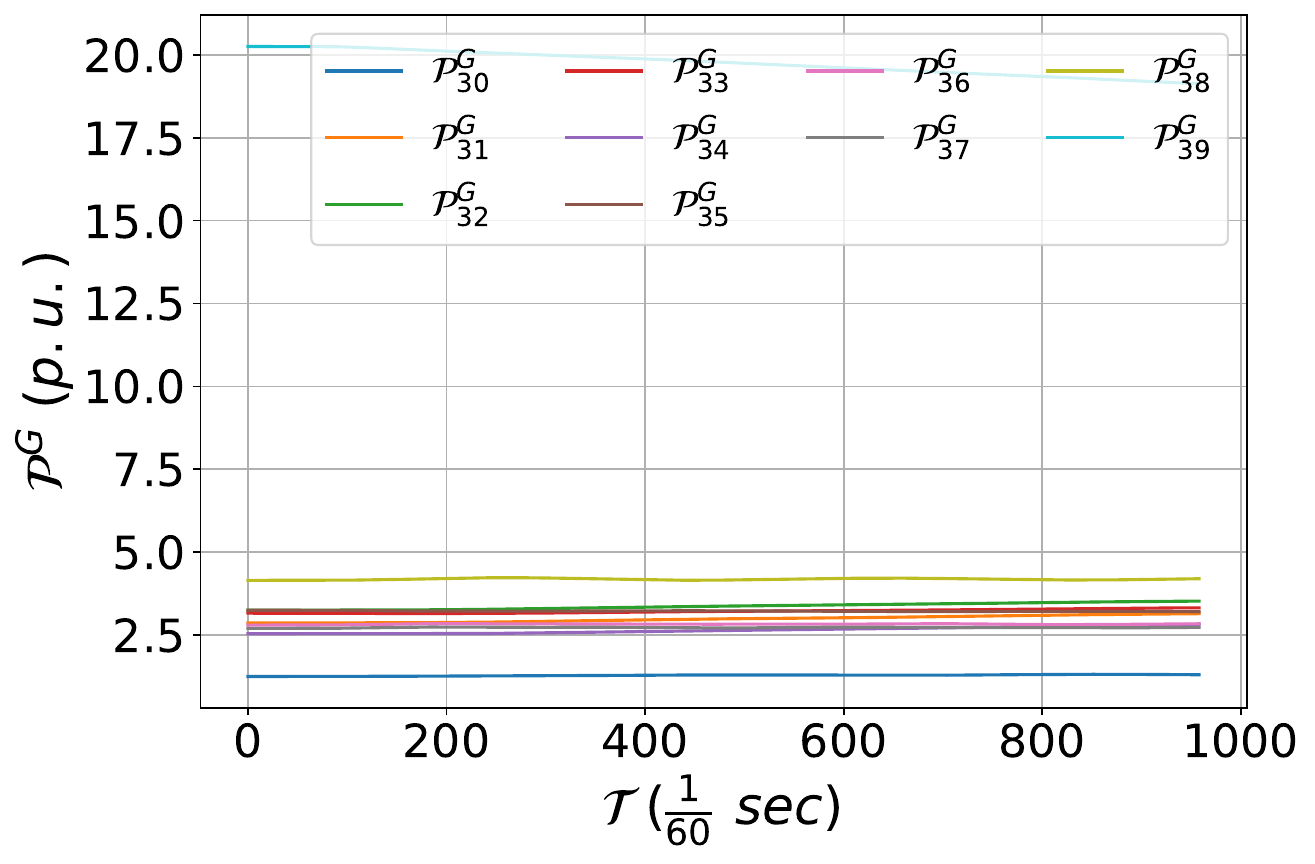}
        }
        \hspace{-10pt}
        \subfigure[]
         {
        \label{case-study-2-pr}
            \includegraphics[width=0.49\columnwidth]{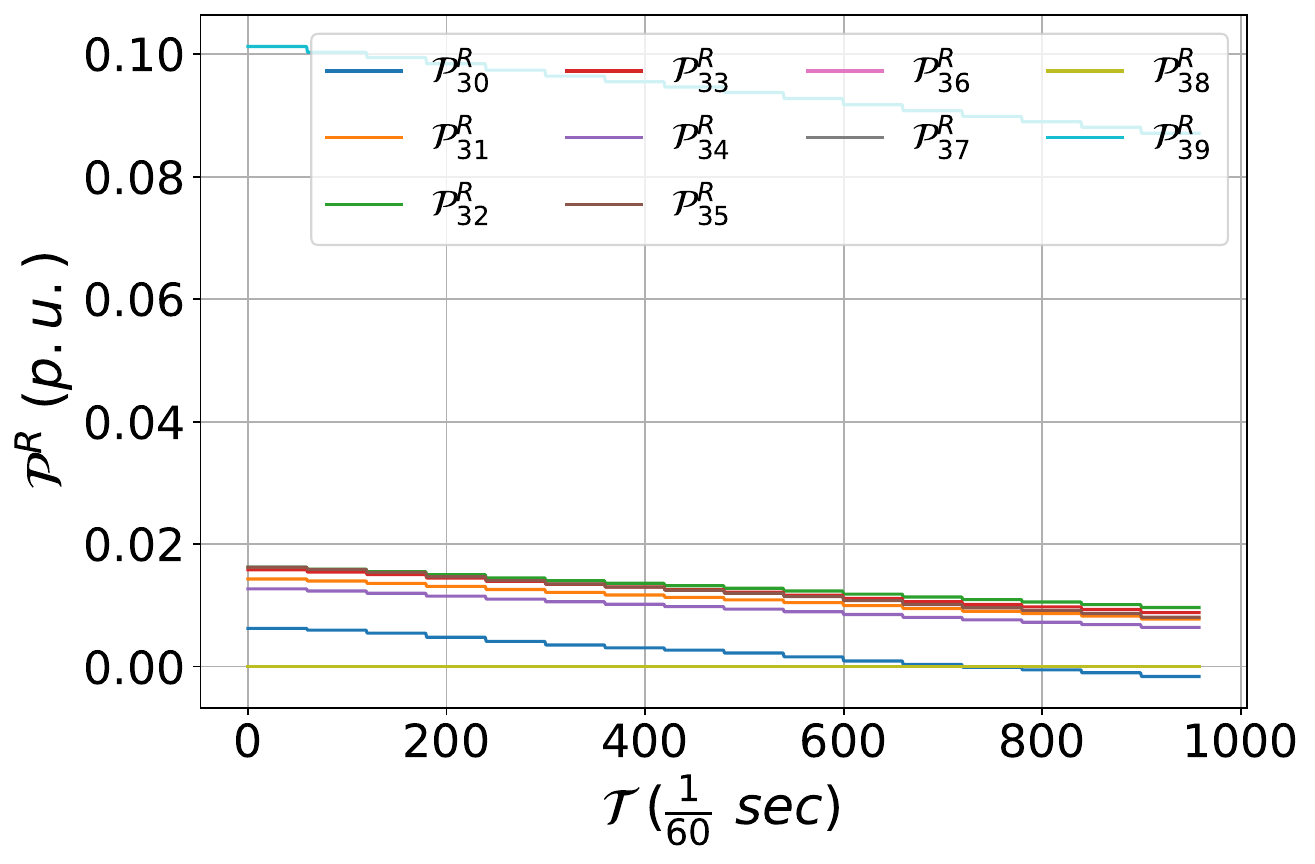}
        }
        \hspace{-10pt}
        \subfigure[]
         {
        \label{case-study-2-f}
            \includegraphics[width=0.48\columnwidth]{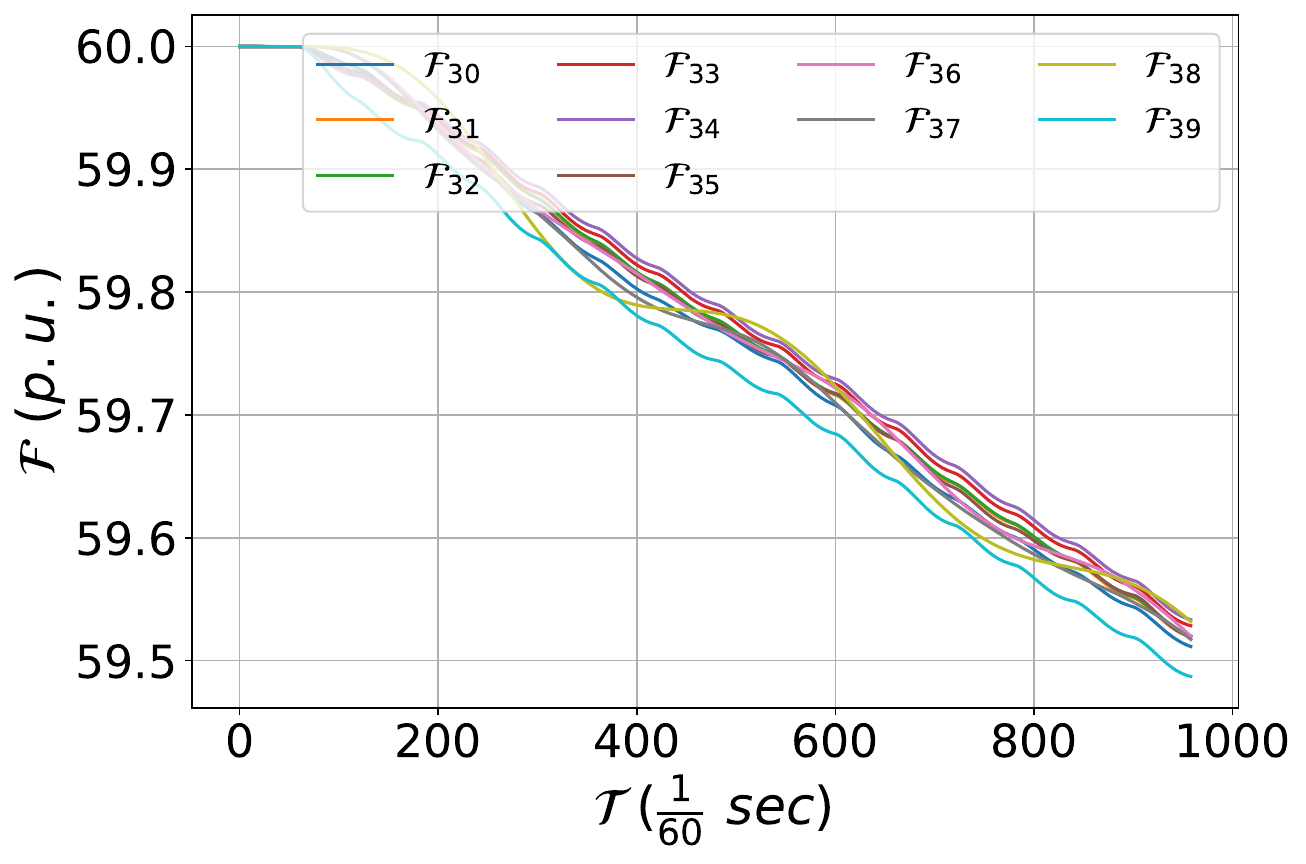}
        }
    \end{center}
    \vspace{-15pt}
    \caption{\small Demonstrating the (a) attacked load measurements (p.u.) from buses and (b) generated active power (p.u.), (c) reference setpoint (p.u.), and (e) frequency (Hz) of different SG generators for UF relay attack in the presence of rules-based BDD.}
    \label{fig:case-study-2}
    \vspace{-6pt}
\end{figure*}
%%%%%%%%%%%%

%%%%%%%%%%%%%%%%%% Case Study 3 %%%%%%%%%%%%%%%%%%%%
%%%%%%%%%%%%%%%%%%%%%%%%%%%%%%%%%%%%%%%%%%%%%%
\begin{figure*}[!t]
    \begin{center}
        \subfigure[]
        {
        \label{case-study-3-attacked-load}
            \includegraphics[width=0.46\columnwidth]{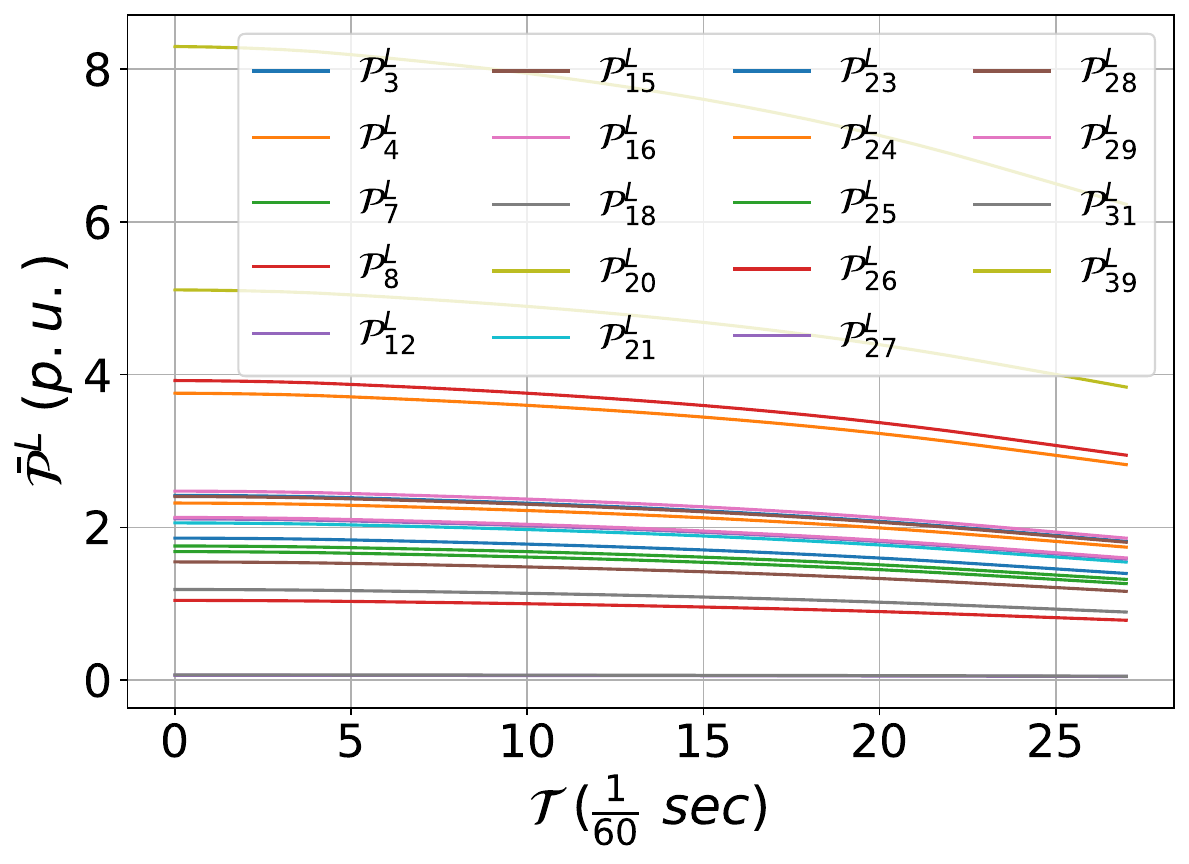}
        }
        \hspace{-10pt}
        \subfigure[]
         {
        \label{case-study-3-pg}
            \includegraphics[width=0.5\columnwidth]{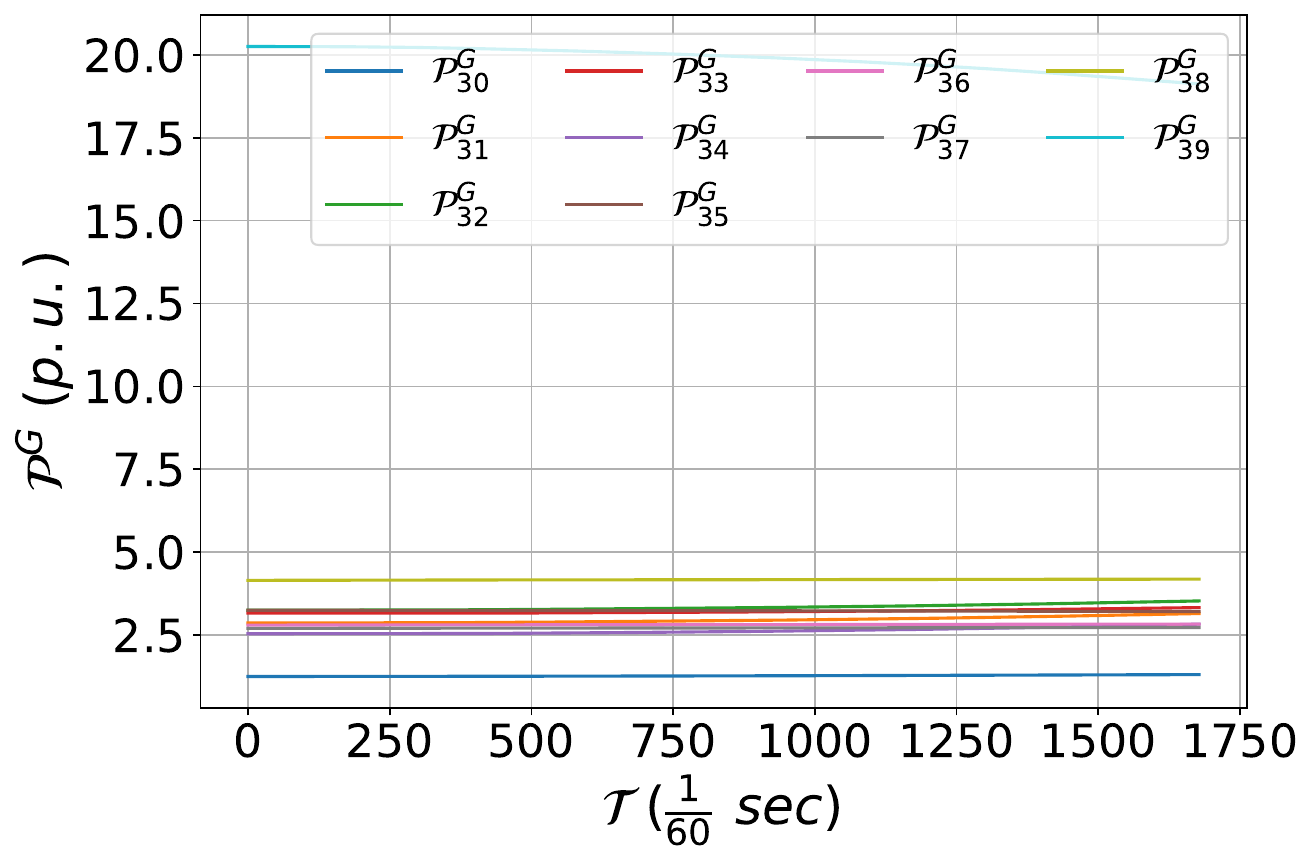}
        }
        \hspace{-10pt}
        \subfigure[]
         {
        \label{case-study-3-pr}
            \includegraphics[width=0.5\columnwidth]{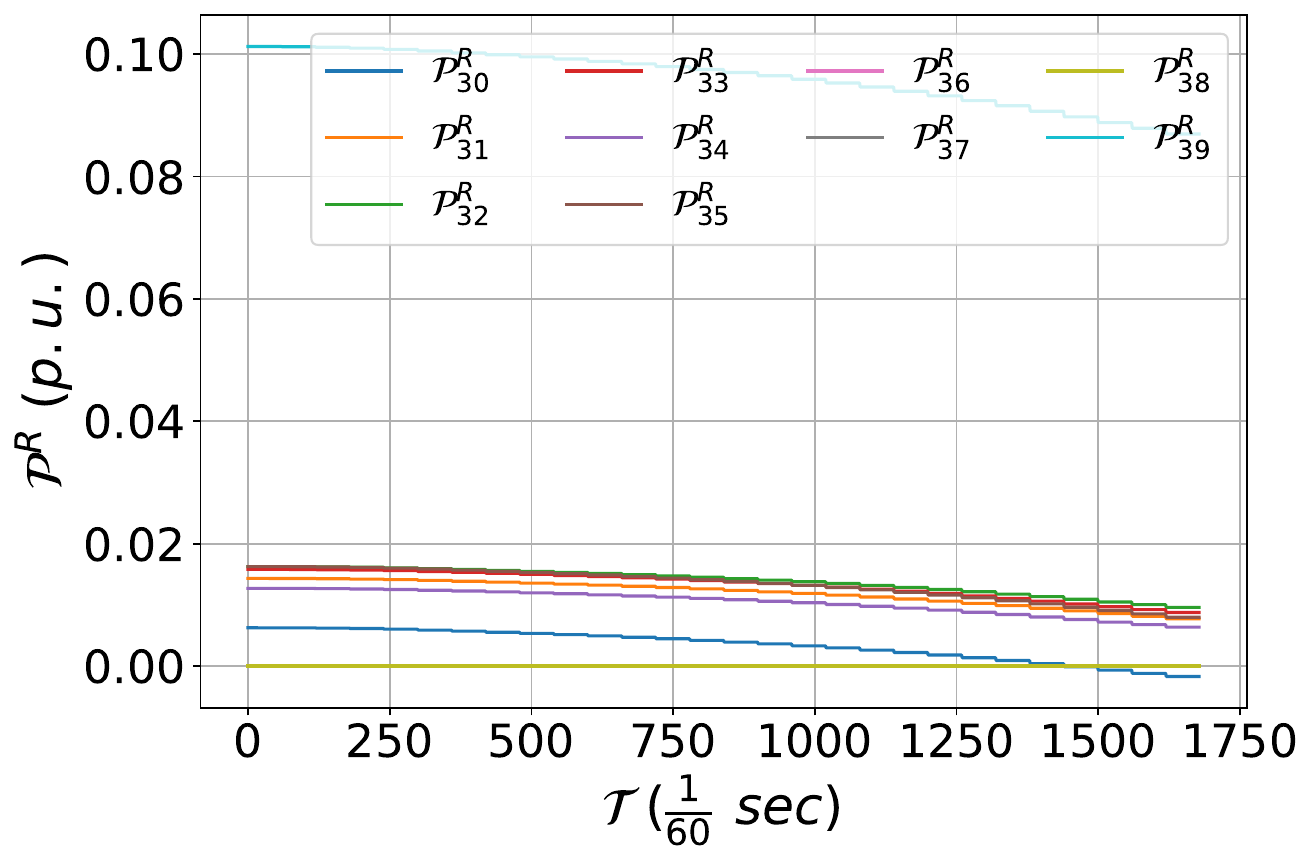}
        }
        \hspace{-10pt}
        \subfigure[]
         {
        \label{case-study-3-f}
            \includegraphics[width=0.49\columnwidth]{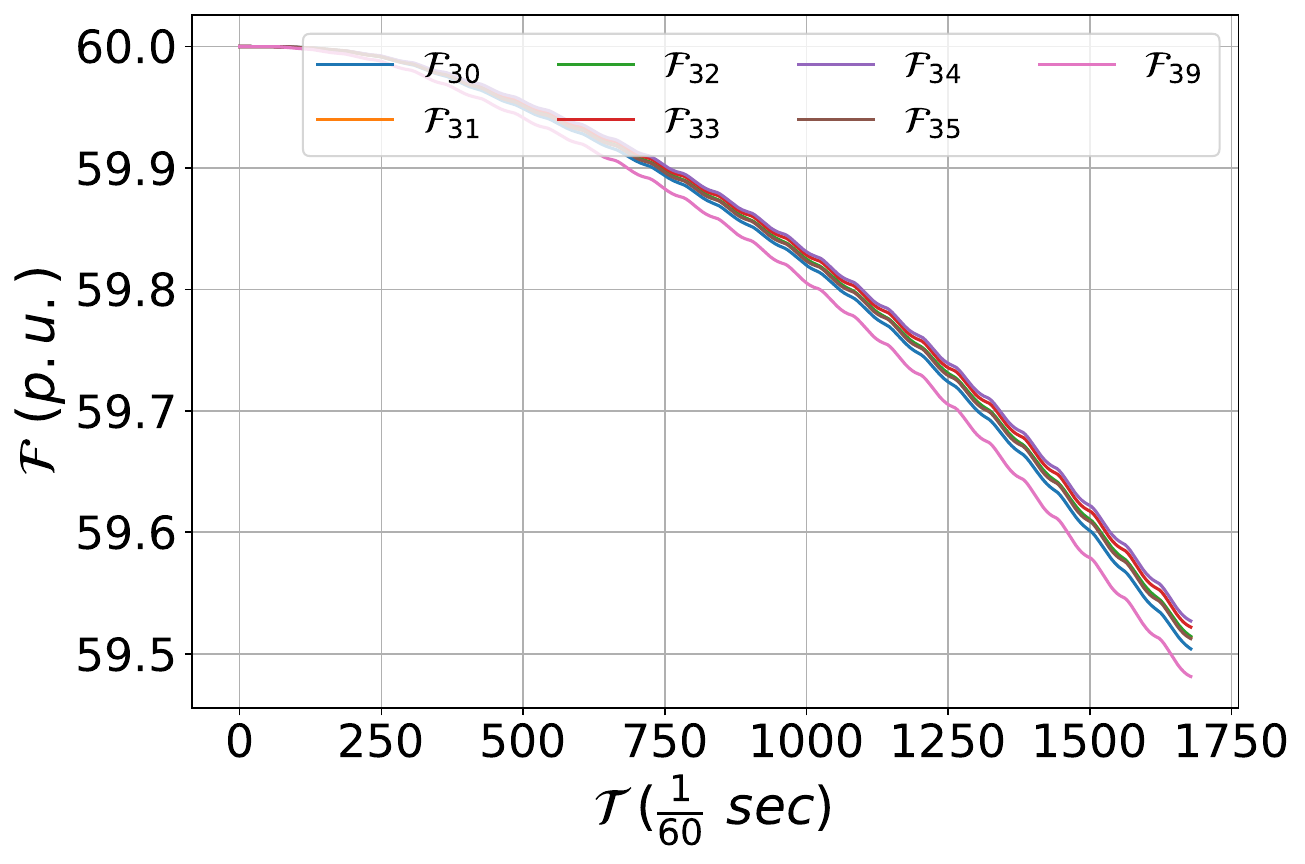}
        }
    \end{center}
    \vspace{-15pt}
    \caption{\small Demonstrating the (a) attacked load measurements (p.u.) from buses and (b) generated active power (p.u.), (c) reference setpoint (p.u.), and (e) frequency (Hz) of different SG generators for UF relay attack in the presence of ML-based ADM.}
    \label{fig:case-study-3}
    %\vspace{-6pt}
\end{figure*}
%%%%%%%%%%%%

%%%%%%%%%%%%%%%%%% Case Study 4 %%%%%%%%%%%%%%%%%%%%
%%%%%%%%%%%%%%%%%%%%%%%%%%%%%%%%%%%%%%%%%%%%%%
\begin{figure*}[!t]
    \begin{center}
        \subfigure[]
        {
        \label{case-study-4-attacked-load}
            \includegraphics[width=0.45\columnwidth]{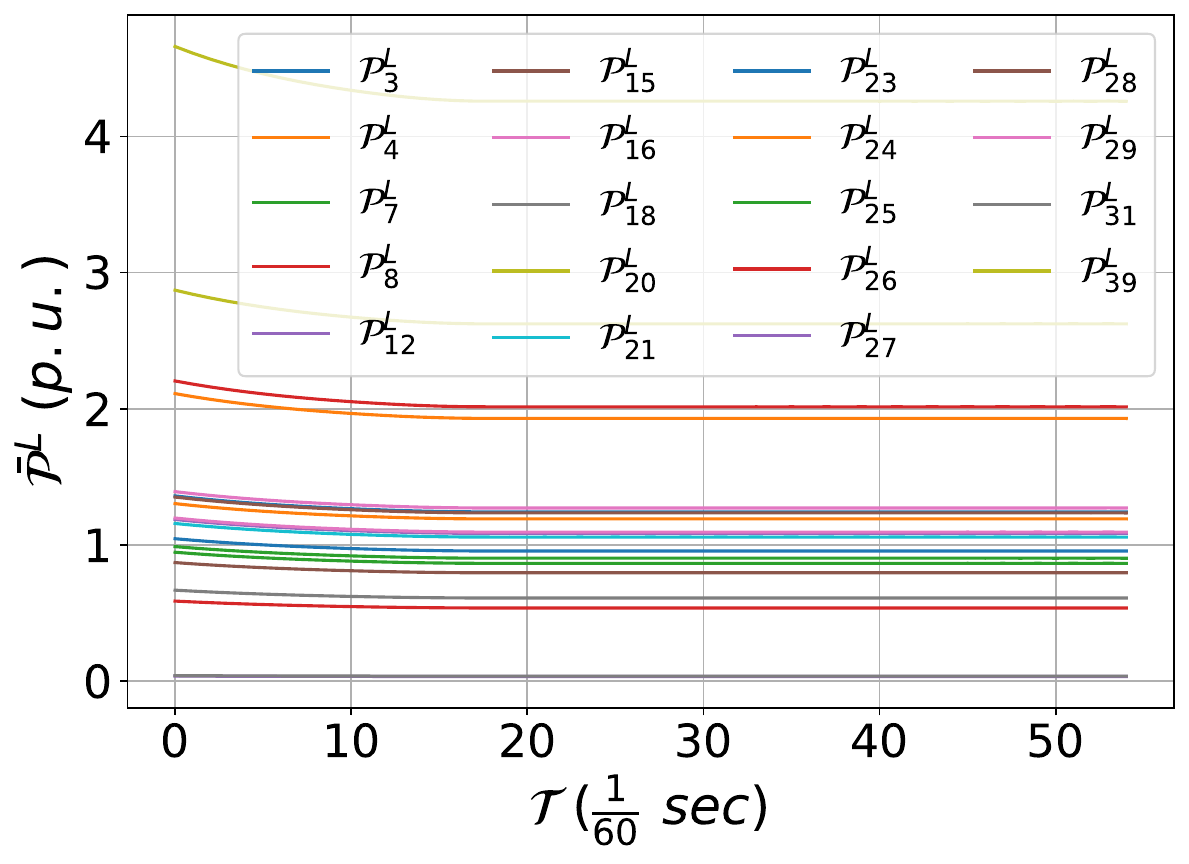}
        }
        \hspace{-10pt}
        \subfigure[]
         {
        \label{case-study-4-pg}
            \includegraphics[width=0.45\columnwidth]{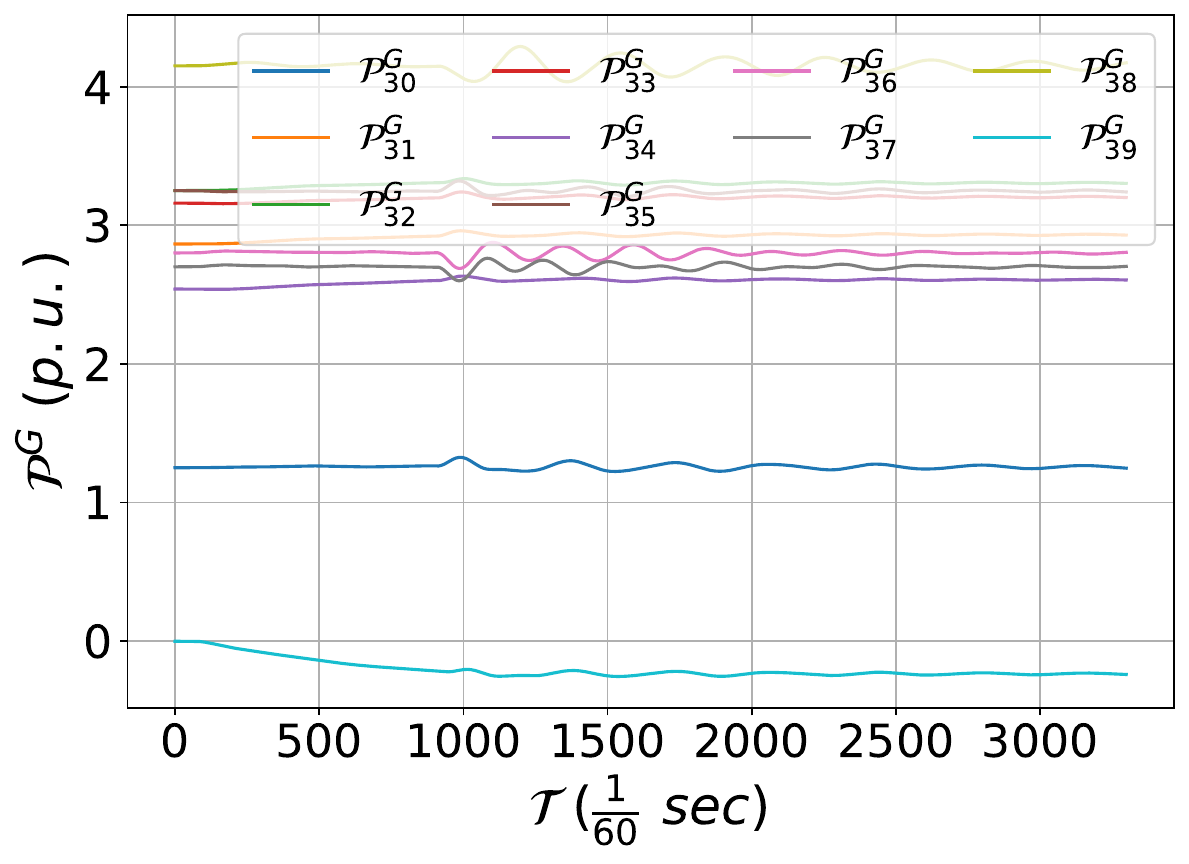}
        }
        \hspace{-10pt}
        \subfigure[]
         {
        \label{case-study-4-pr}
            \includegraphics[width=0.51\columnwidth]{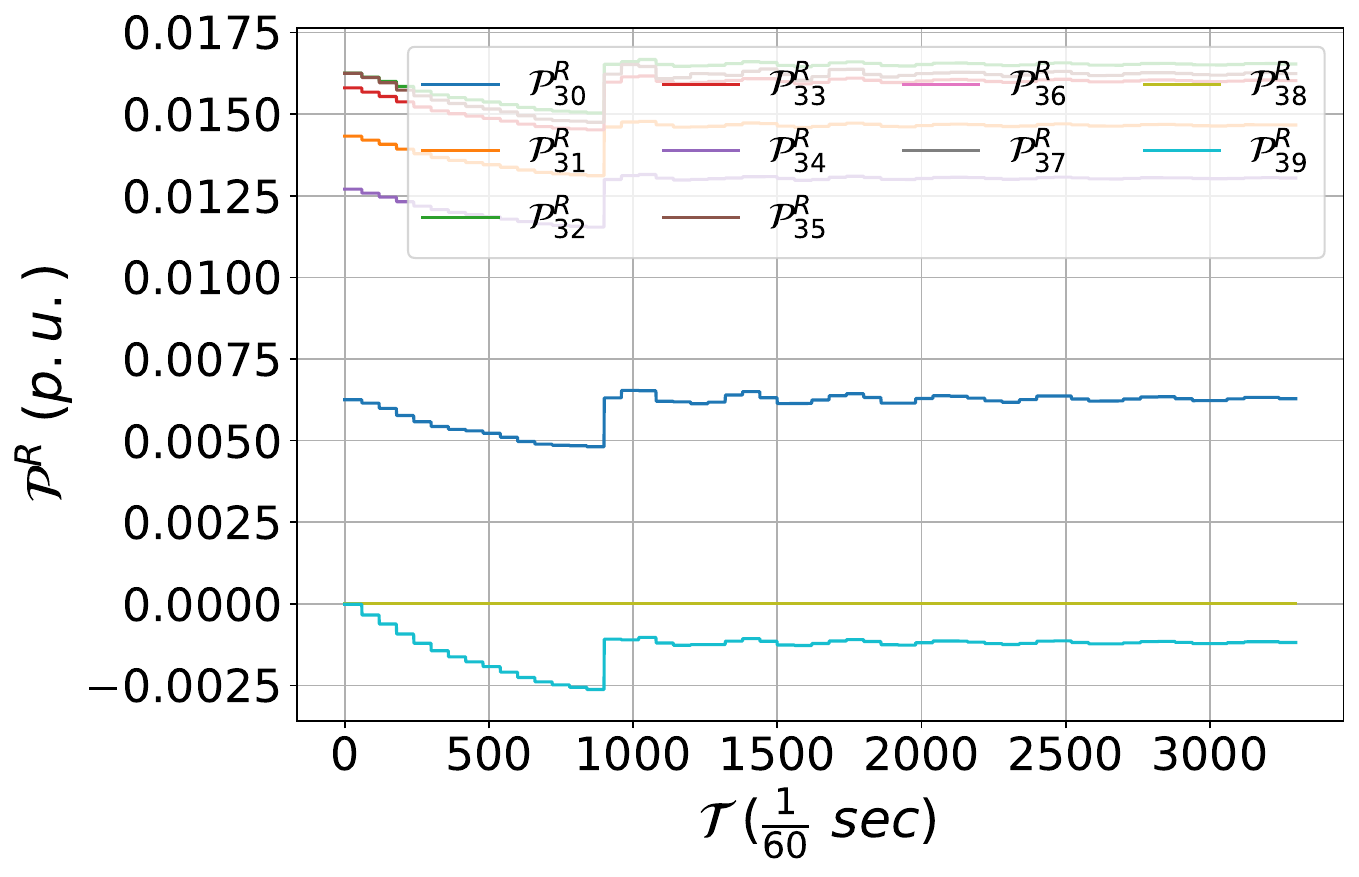}
        }
        \hspace{-10pt}
        \subfigure[]
         {
        \label{case-study-4-f}
            \includegraphics[width=0.48\columnwidth]{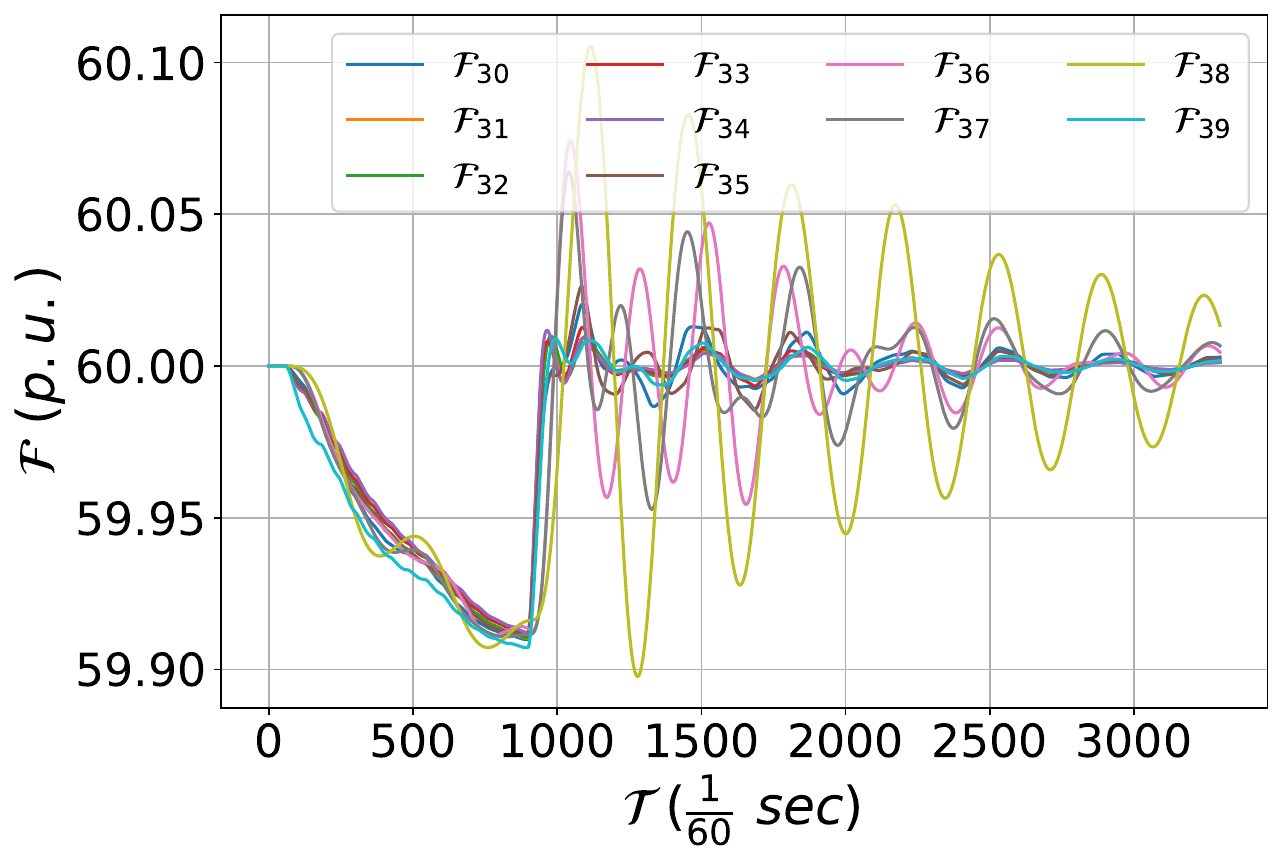}
        }
        \hspace{-10pt}
    \end{center}
    \vspace{-15pt}
    \caption{\small Demonstrating the (a) attacked load measurements (p.u.) of buses and (b) generated active power (p.u.), (c) mechanical power (p.u.), and (d) reference setpoints (p.u.) of different SG generators for discontinued UF attack.}
    \label{fig:case-study-4}
    %\vspace{-6pt}
\end{figure*}
%%%%%%%%%%%%

This section provides different case studies demonstrating the working principle of the \framework~analytics. Here, we demonstrate how the analytics operates for benign load measurements and can trigger UF relays with rules-based BDD and ML-based ADM. In the case of ML-based ADM, we illustrate that even when a stealthy FDI attack can no longer alter the load measurements, and although the attacked load measurement differs from the benign load measurement, the presence of the primary controller maintains the synchronicity of the generator. Furthermore, we demonstrate the validity of our formal modeling using a case study that showcases how discontinuing the attack restores the frequency to its nominal value. Our analysis uses data from the GEFCom2014 load forecasting dataset, which collected hourly data from the New England IEEE 39-Bus System. To handle the dataset, we employed a curve-fitting approach for data imputation. The load measurements show very minor deviations between consecutive readings. We treat the 10-minute intervals between load measurements in the GEFCom2014 dataset to replicate transient behavior as successive measurements. Both case studies and evaluations utilize this processed dataset.

\vspace{-10pt}
\subsection{Case Study on Benign Scenario}

When the load in the system changes, the LFC dispatches the adjusted reference setpoint of the generators. Figure~\ref{fig:case-study-1} illustrates the benign response of the system. The analysis involves a deliberate, benign alteration of load measurements for the initial 3000 timeslots, equivalent to 50 seconds. The fluctuation of load measurement can be seen from Figure~\ref{case-study-1-load}. Although there are 39 buses, we only present measurements for 29 buses because the load measurements for the remaining buses are nearly 0 MW. We can also observe from Figures~\ref{case-study-1-pg}, \ref{case-study-1-pr} that the active power of the generators and their reference setpoints show deviations during the time interval when load measurements are altered. The mechanical power is omitted from the figure since it closely follows the pattern of active generating power. The impact of this experimental setup is evident in the observable oscillations depicted in Figure~\ref{case-study-1-f}, which show the frequency oscillation of the individual generators. It seems that despite the initial oscillations, the frequency of the generators is returning to the nominal frequency. The stabilization is attributed to the LFC's dynamic adjustment of reference power.  

\vspace{-10pt}
\subsection{Case Study on Attacked Scenario with Rules-based BDD}
In this case study, we investigate the system's behavior under the influence of an FDI attack designed to activate the UF relays of generators. A BDD from TIFS'23 is considered to detect anomalies from LFC load measurements. During this analysis, we maintained benign load measurements at a constant level. For launching an attack, the load measurements are altered to make the grid unstable, as shown in Figure~\ref{case-study-2-attacked-load}. Hence, the active power and generator reference powers also got impacted, as demonstrated in Figures~\ref{case-study-2-pg} and~\ref{case-study-2-pr} are impacted. The attack is carried out from the 60th timeslot and is continued until at least one of the generator's frequencies dropped below 59.5 Hz (see Figure~\ref{case-study-2-f}). It was observed that after 953 timeslots, the frequency in the generator connected to the 39th bus employing the load frequency controller with rules-based BDD dropped to 59.5 Hz, triggering the UF relay. This analysis restricted the attacker's access to the load measurement to 15 buses.

\vspace{-10pt}
\subsection{Case Study on Attacked Scenario with ML-based ADM}
The purpose of this case study is to compare the impacts of ML-based ADM with rules-based BDD. We maintain the same benign load measurements, external conditions, and attack goals as in the previous case study. However, from Figure~\ref{case-study-3-attacked-load}-\ref{case-study-3-pr}, the alterations in load measurements and the corresponding impacts on generating active power and reference power are less drastic compared to the previous case. Consequently, the affected frequency, as shown in Figure~\ref{case-study-3-f}, takes longer to trigger the UF relay. Specifically, the UF relay was activated at the 1578th timeslot, which is 65\% more timeslots (or 13 Load Frequency Control cycles) than in the attack conducted with the rules-based BDD. This demonstrates that the ML-based ADM significantly enhances attack resiliency.

%%%%%%%%%%%%%%%%%% Validation %%%%%%%%%%%%%%%%%%%%
%%%%%%%%%%%%%%%%%%%%%%%%%%%%%%%%%%%%%%%%%%%%%%
\begin{figure*}[!t]
    \begin{center}
        \subfigure[]
        {
        \label{val-1-f}
            \includegraphics[width=0.48\columnwidth]{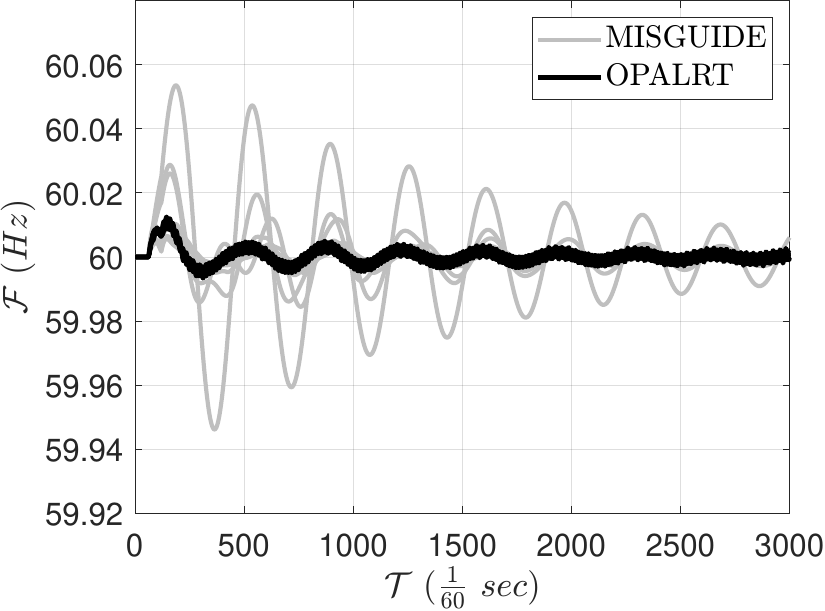}
        }
        \subfigure[]
         {
        \label{val-2-f}
            \includegraphics[width=0.46\columnwidth]{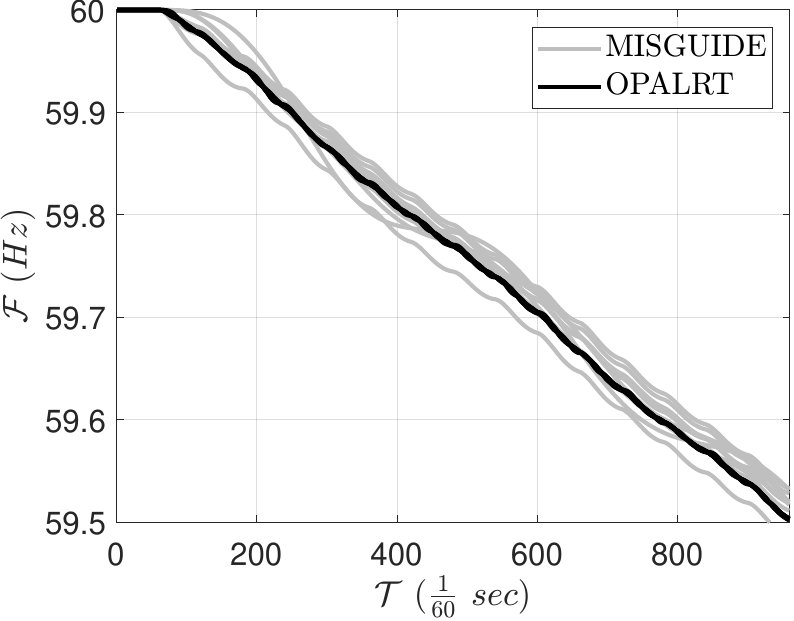}
        }
        \subfigure[]
        {
        \label{val-3-f}
            \includegraphics[width=0.46\columnwidth]{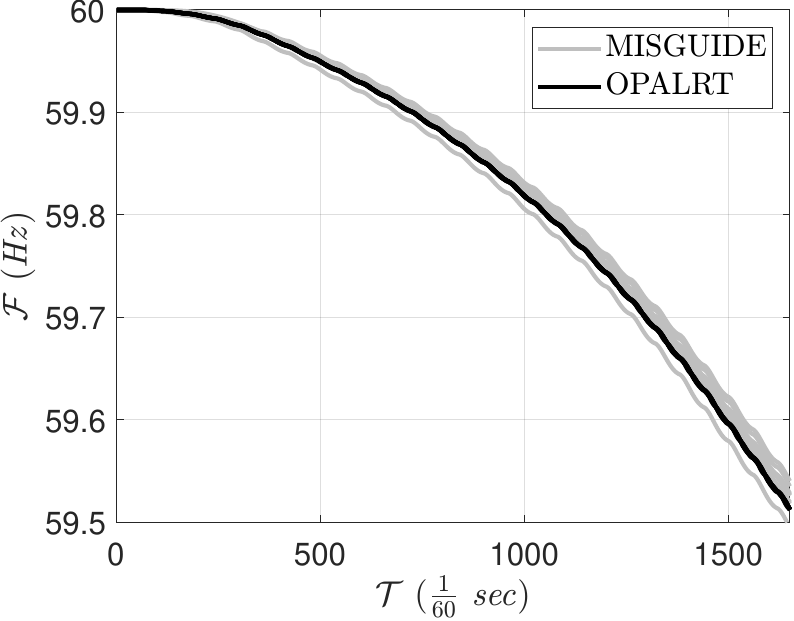}
        }
        \subfigure[]
         {
        \label{val-4-f}
            \includegraphics[width=0.46\columnwidth]{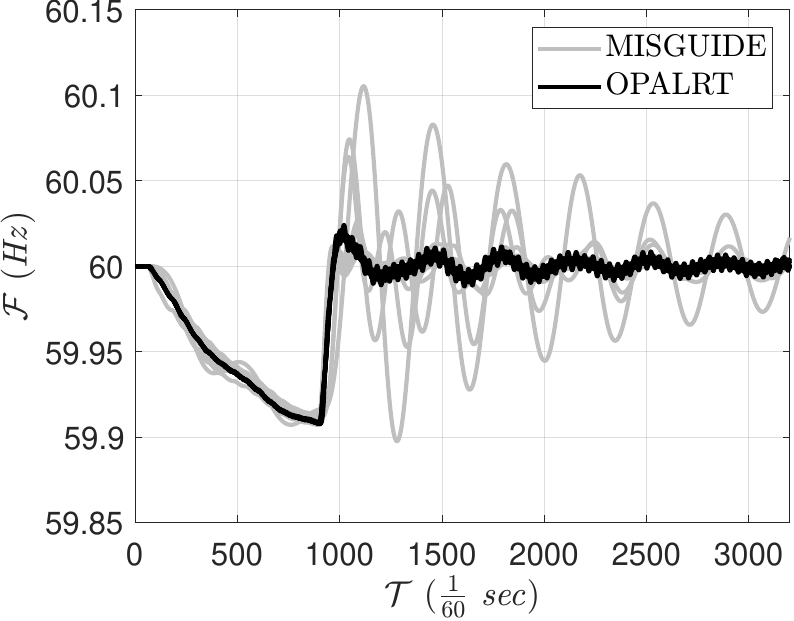}
        }
    \end{center}
    \vspace{-15pt}
    \caption{\small Demonstrating the comparative frequency measurements between ~\framework and OPALRT for (a) case study 1, (b) case study 2, (c) case study 3, and (d) case study 4.}
    \label{fig:validation}
    \vspace{-6pt}
\end{figure*}

%%%%%%%%%%%%%%%%%% Evaluation 1 %%%%%%%%%%%%%%%%%%%%
%%%%%%%%%%%%%%%%%%%%%%%%%%%%%%%%%%%%%%%%%%%%%%
\begin{figure*}[!t]
    \begin{center}
        \subfigure[]
        {
        \label{eval-1a-attacked-load}
            \includegraphics[width=0.46\columnwidth]{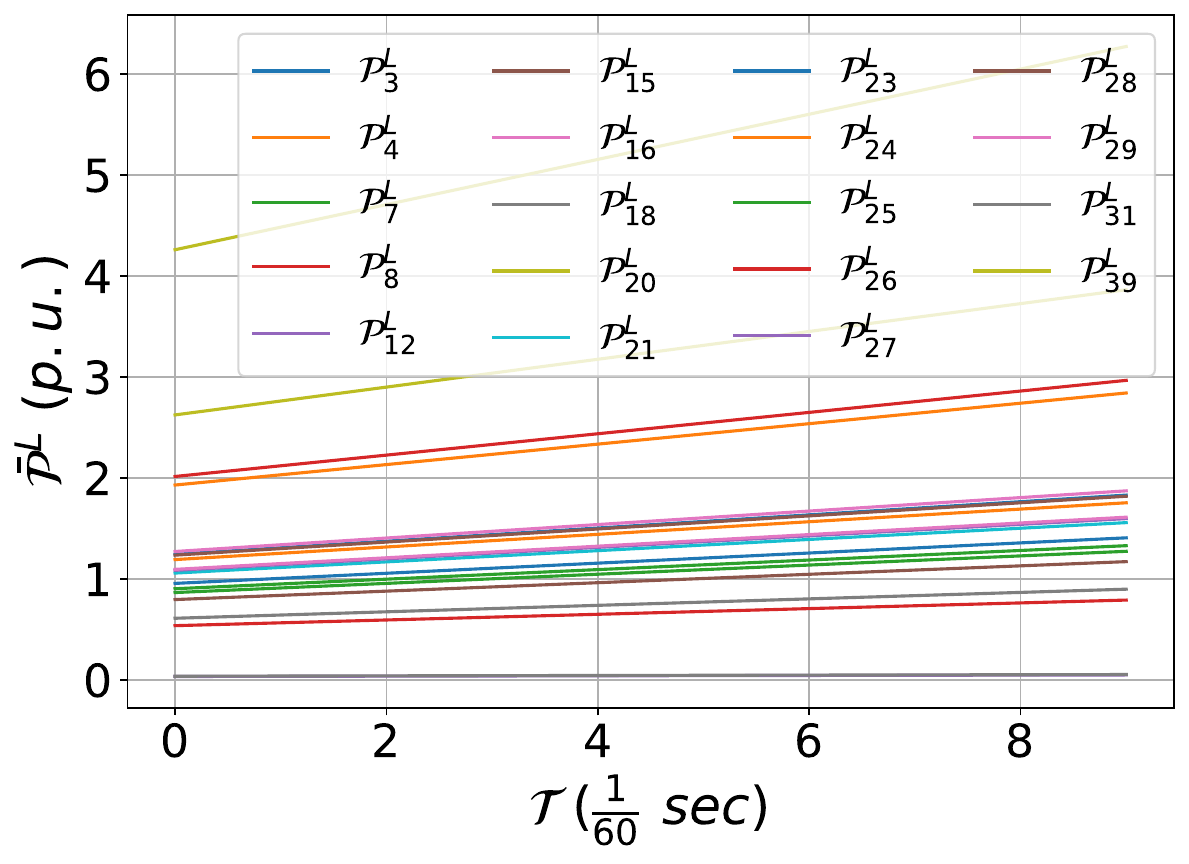}
        }
        \subfigure[]
         {
        \label{eval-1a-f}
            \includegraphics[width=0.48\columnwidth]{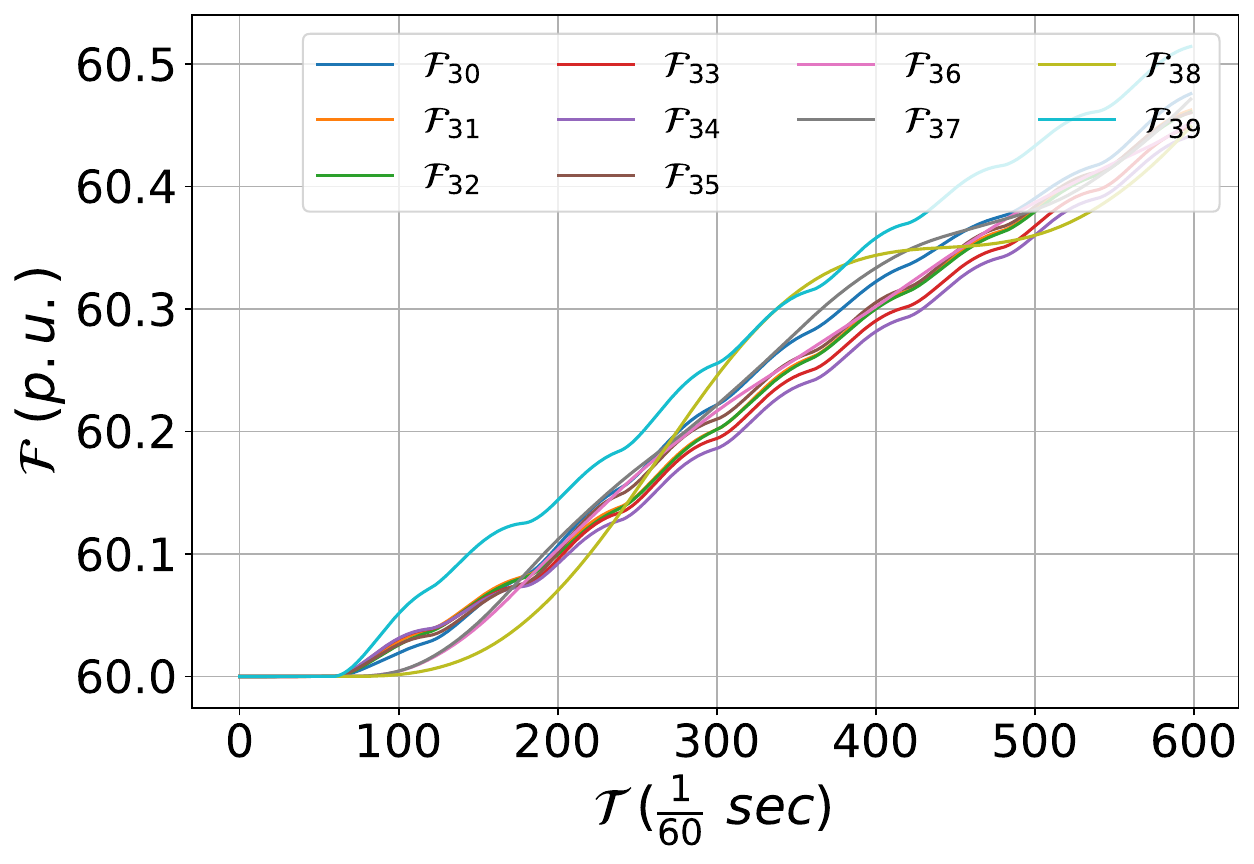}
        }
        \subfigure[]
        {
        \label{eval-1b-attacked-load}
            \includegraphics[width=0.46\columnwidth]{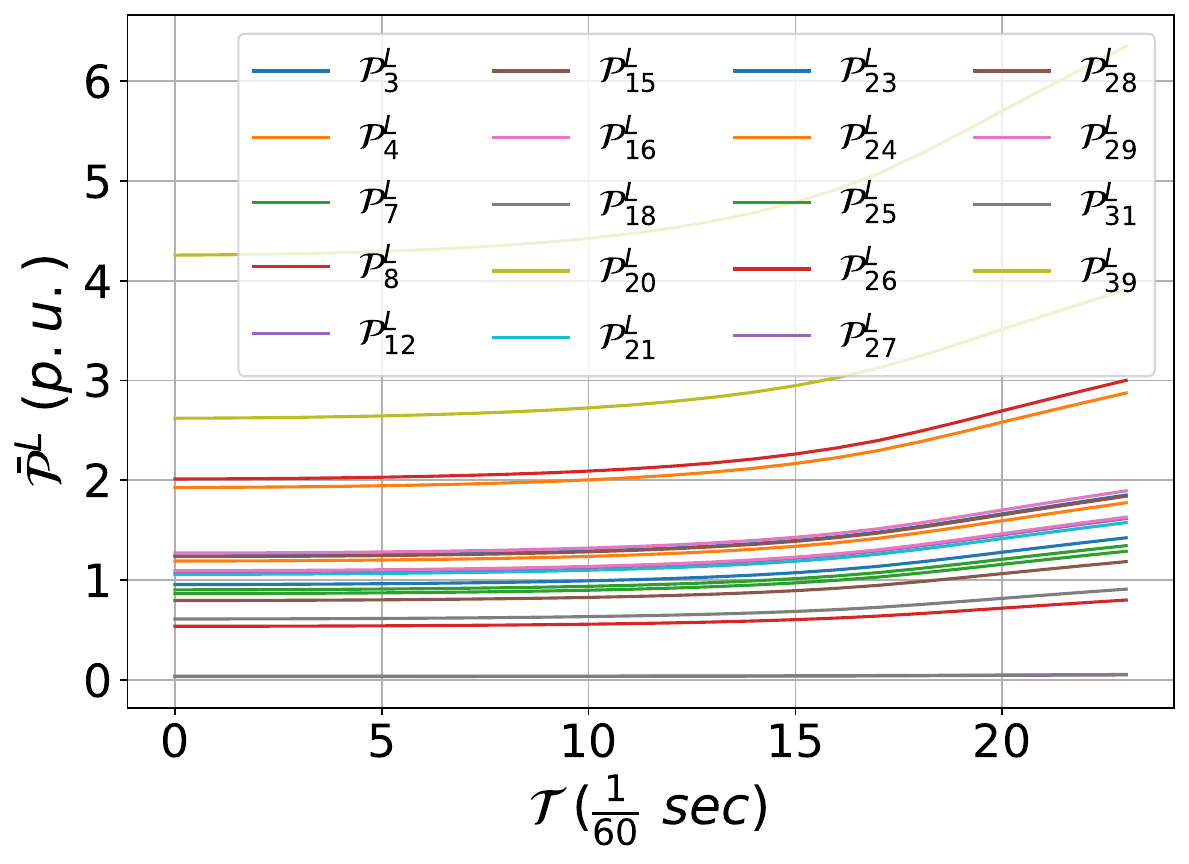}
        }
        \subfigure[]
         {
        \label{eval-1b-f}
            \includegraphics[width=0.48\columnwidth]{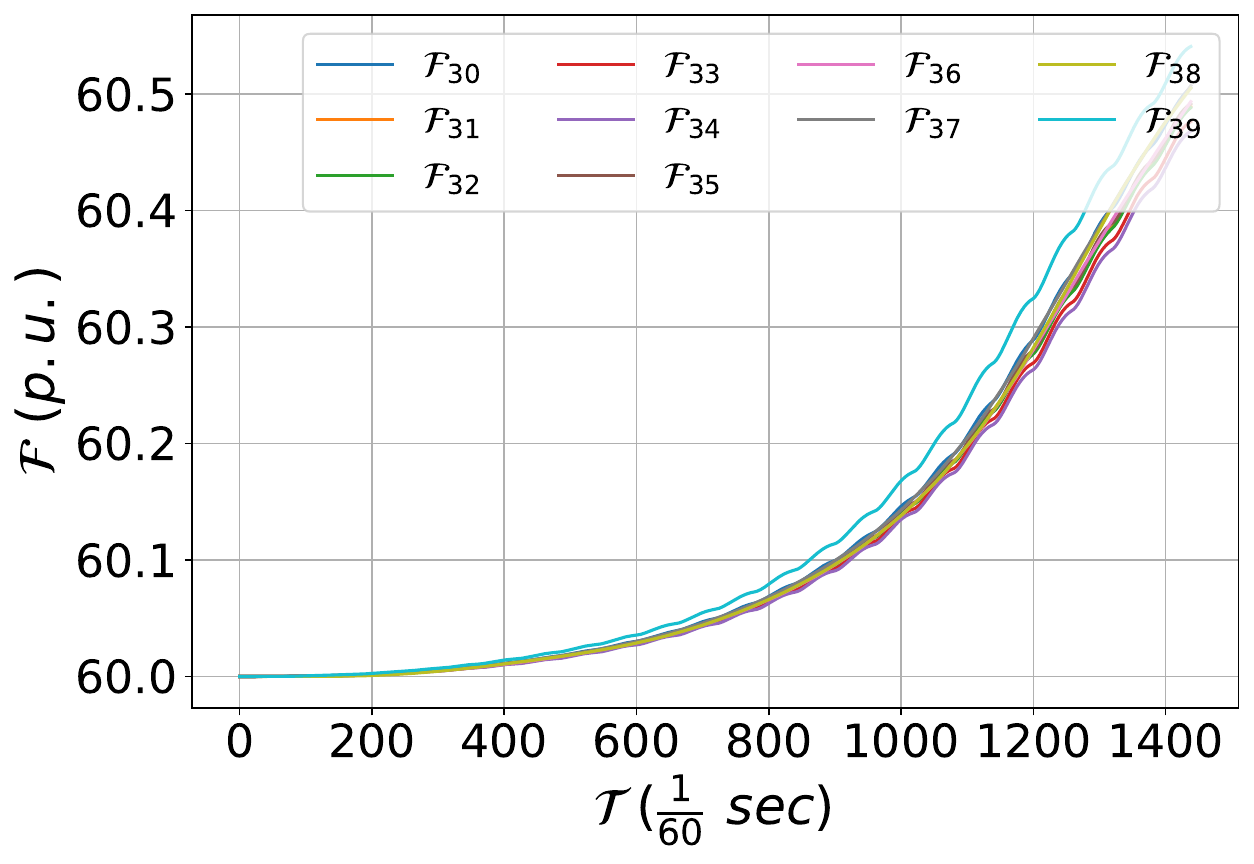}
        }
    \end{center}
    \vspace{-15pt}
    \caption{\small Demonstrating the (a) attacked load measurements (p.u.), (b) frequency (Hz) in the presence of rules-based BDD, and (c) attacked load measurements (p.u.), (d) frequency (Hz) in the presence of ML-based ADM of different SG generators for OF relay attack.}
    \label{fig:adm-attack-impact}
    \vspace{-6pt}
\end{figure*}
%%%%%%%%%%%%

%
\subsection{Case Study on Discontinuation of Attack}
We validate our formal modeling through a case study that demonstrates how discontinuing an attack restores the frequency to its nominal value. In this case study, FDI attacks were conducted on UF relays for 16 LFC cycles before being stopped. During the attack, benign measurements were kept constant, similar to previous case studies. However, the attack affected the load measurements perceived during the LFC cycles, as shown in Figure~\ref{case-study-4-attacked-load}. The impact on load measurements influenced the active generation and reference powers, as illustrated in Figures~\ref{case-study-4-pg} and~\ref{case-study-4-pr}. After discontinuing the attack, the frequency measurements of the generators returned to the nominal frequency following some oscillations, as depicted in Figure~\ref{case-study-4-f}. The recovery supports the accuracy and correctness of our formal modeling.

\vspace{-12pt}
\subsection{Validation}
Here, we validate the attack vectors identified by \framework~using Ephasorsim of OPAL-RT, a real-time simulator. For the case studies, the maximum frequency deviations are 0.0485 Hz, 0.0323 Hz, 0.0293 Hz, and 0.1088 Hz, respectively. The deviations can be attributed to 1) linearization for power system dynamics modeling, 2) omission of states in generator modeling, and 3) exclusion of voltage control components in the dynamics modeling. \framework~considers only two states for generator modeling - angular frequency and rotor angle. Voltage control components like automatic voltage regulators and power system stabilizers have a loosely coupled relationship with frequency control, which is not considered in the proposed framework. Although slight deviations exist between the modeling and simulation results, the deviations are substantially small to be ignored. Figure~\ref{fig:validation} shows that the impacted \framework~frequency measurements closely follow the OPAL-RT frequency measurements for all the case studies. From Figure~\ref{val-1-f}, it might appear that the OPAL-RT results significantly deviate from those of the \framework~considered controller. However, this discrepancy is due to the time scale being much smaller compared to the other figures.
%%%%%%%%%%%%%%%%%%%%%%%%%%%%%
%%%%%%%%%%%%%%%%%%%%%%%%%%%%%%%%%%%%%%%%%%
%%%%%%%%%%%%%%%%%%%%%%%%%%%%%%%%%%%%%%%%%%
\section{Evaluation}
\label{sec:evaluation}

We evaluate the feasibility and effectiveness of our proposed framework, including its findings on SG FDI attack vectors and their impacts. \framework~is assessed on the following research questions (RQs). 

\textbf{RQ1:} What is the impact of the MISGUIDE-generated attack vectors? (Section~\ref{subsec:eval_adm})

\textbf{RQ2:} What are the framework's findings in assessing the impact of proposed attacks considering various levels of adversarial capability? (Section~\ref{subsec:eval_adversarial-attributes})

\textbf{RQ3:} What are the framework's findings in assessing the system's resiliency? (Section~\ref{subsec:eval_resiliency})

\textbf{RQ4:} How scalable is the proposed framework? (Section~\ref{subsec:eval_scalability})

%%%%%%%%%%%%%%%%%%%%%%%%%%%%%
%%%%%%%%%%%%%%%%%%%%%%%%%%%%%
\subsection{Evaluation of Anomaly Detection Model}
\label{subsec:eval_adm}
The prevalence of ML-based ADM increases the robustness of the system. Here, we analyze the impact of ADM compared to BDD rules-based attack goal achieving time. The analysis of the UF relay-triggering attack was demonstrated in case study 1. Here, we are showing the analysis for an OF relay attack. Figure~\ref{fig:adm-attack-impact} shows the attack impact through frequency measurements of different generators. During the analysis with rules-based BDD, we maintained benign load measurements at a constant level. From Figure~\ref{eval-1a-attacked-load}, we can see that the attack was carried out from the 60th timeslot and continued until the end of the analysis (i.e., the 600th timeslot) in the presence of rules-based BDD. It was observed that after 9 seconds, the frequency of generator bus 38 goes above 60.5 Hz, which trips the generator associated with the bus as shown in Figure~\ref{eval-1a-f}. Evading ML-based ADM, the load measurement alteration is more restricted than the rules-based BDD case (see Figure~\ref{eval-1b-attacked-load}). It takes three times more time than rules-based BDD to isolate a generator from the system, as seen in Figure~\ref{eval-1b-f}. The OF relay triggering takes slightly less time than the UF relay triggering. The reason is that the BDD and ADM threshold for an increase in load measurements is more flexible than the decrease in load measurements, as learned from the benign load patterns.

%%%%%%%%%%%%%%%%%%%%%%%%%%%%%
%%%%%%%%%%%%%%%%%%%%%%%%%%%%%
\subsection{Evaluation of Attack Impact with Adversarial Attributes}
\label{subsec:eval_adversarial-attributes}
Here, we evaluate the SG based on different adversarial attributes in this section as shown in Table~\ref{tab:attacker's-accessibility}. This analysis focuses on the regular load pattern starting from day 4 of the dataset. The table shows that attackers with access to fewer measurements require more time to achieve their attack goal. For instance, with a rules-based BDD and access to 15 measurements, triggering an OF relay takes 11 LFC cycles. However, if attackers can only access 5 load measurements, it takes 25 LFC cycles to trigger an OF relay. Another observation is that ADM can better accommodate increases in load measurements compared to decreases. As a result, the OF relay is triggered more quickly than the UF relay. For example, the UF relay takes 27 LFC cycles to activate, while the OF relay can be triggered in 21 cycles in the case of ML-based ADM. It is important to note that the OF relay is more critical because its activation can isolate a generator from the system, leading to significant recovery time and cost. Additionally, isolating a generator can have cascading impacts on the overall system.

%%%%%%%%%%%%%%%%%%%%%%%%%%%%%% 
\begin{table}[!t] 
\scriptsize 
\centering
\label{tab:attacker's-accessibility}
\caption{Attack Impact with Various Attacker's Accessibility} 
\vspace{-6pt}
\begin{tabular}{|l|l|c|c|} \hline Defense Type & Attack Goal & \begin{tabular}[c]{@{}c@{}}Access to Load \\ Measurements\end{tabular} & \begin{tabular}[c]{@{}c@{}}Number of Timeslots (LFC \\Cycles) to Attain Attack Goal \end{tabular} \\ \hline 
\multirow{8}{*}{\begin{tabular}[c]{@{}l@{}}Rules-Based\\ BDD\end{tabular}} & 
\multirow{4}{*}{UF} & 5 & 2013 (33)\\ 
\cline{3-4} & & 10 & 1089 (18)\\ 
\cline{3-4} & & 15 & 953 (15)\\ 
\cline{3-4} & & 19 & 787 (13)\\ 
\cline{2-4} &
\multirow{4}{*}{OF} & 5 & 1547 (25)\\ 
\cline{3-4} & & 10 & 845 (14)\\ 
\cline{3-4} & & 15 & 693 (11)\\ 
\cline{3-4} & & 19 & 416 (6)\\ \hline 
\multirow{8}{*}{\begin{tabular}[c]{@{}l@{}}ML-based \\ ADM\end{tabular}} & 
\multirow{4}{*}{UF} & 5 & 2823 (47)\\
\cline{3-4} & & 10 & 1633 (27)\\
\cline{3-4} & & 15 & 1301 (21)\\
\cline{3-4} & & 19 & 1225 (20)\\
\cline{2-4} &
\multirow{4}{*}{OF} & 5 & 1987 (33)  \\ 
\cline{3-4} & & 10 & 1279 (21)\\ 
\cline{3-4} & & 15 & 1059 (17)\\ 
\cline{3-4} & & 19 & 963 (16)\\ \hline \end{tabular} 
\vspace{-6pt}
\end{table} 
%%%%%%%%%%%%%%%%%%%%%%%%%%%%%

%%%%%%%%%%%%%%%%%%%%%%%%%%%%%
% %%%%%%%%%%%%%%%%%%%%%%%%%%%%%
% \subsection{Evaluation of the System's Critical Sensor Measurements}
% \label{subsec:eval_critical-measurements}
%

%%%%%%%%%%%%%%%%%%%%%%%%%%%%%
%%%%%%%%%%%%%%%%%%%%%%%%%%%%%
\subsection{Evaluation of the Resiliency Analysis of the System}
\label{subsec:eval_resiliency}

The ADM might not provide complete protection against FDI attacks since attackers with access to ADM parameters can still launch stealthy attacks and evade them. Hence, we further evaluate the threat space of the ADM-Assisted SG system in the presence of ADMs through resiliency analysis, which reveals how well the system protects its critical capabilities from disruption caused by adversarial events and conditions.  
%%%%%%%%%%%%%%%%%%%%%%%%%%%%%%%%%%
\begin{definition}[k-Resilency]
$k-$resiliency refers to the ability of a system, network, or process to withstand and effectively respond to a certain number of $k$ simultaneous failures or attacks without losing its core functionality. In more technical terms, a system is considered "k-resilient if it continues to operate correctly and meet its specified performance and security requirements even when up to k components fail or are compromised. 
\end{definition}

The resilience of a system increases with the value of $k$, where $k$ represents the number of components, such as measurements, that the system can withstand before its functionality is compromised. In this context, $k$ refers to the number of measurements the system can handle under attack conditions. Figure~\ref{tab:resiliency} illustrates the $k$-resiliency of the SG system being analyzed. The figure shows how the system maintains its resilience when an attack is sustained over an extended duration. The analysis indicates that the ML-based ADM demonstrates better resilience than the rule-based BDD system. For instance, an attacker must have access to at least 12 sensor measurements to trigger an OF relay while bypassing the BDD rules successfully. In contrast, even if an attacker can access all 19 measurements, the ML-based ADM effectively prevents them from circumventing its detection capabilities. In this evaluation, we determine that the system's resilience is considered adequate if it can maintain its critical functions with access to at least 11 measurements.

%%%%%%%%%%%%%%%%%%%%%%%%%%%%%%
\begin{table}[!t]
\scriptsize
\centering
\label{tab:resiliency}
\caption{Resiliency Analysis of the System}
\vspace{-6pt}
\begin{tabular}{|p{2cm}|p{1.3cm}|p{2.2cm}|p{1.5cm}|}
\hline 
Type of Defense & Attack Goal & \begin{tabular}[c]{@{}c@{}}Access to Timeslots\end{tabular} & k-Resiliency 
\\ \hline 
\multirow{6}{*}{Rules-Based BDD} 
& \multirow{3}{*}{UF} & 1620 & 11 \\ 
\cline{3-4} & & 1500 & 8 \\ 
\cline{3-4} & & 1380 & 6 \\ 
\cline{2-4} 
& \multirow{3}{*}{OF} & 1260 & 12
\\ \cline{3-4} & & 1140 & 6
\\ \cline{3-4} & & 1020 & 4\\ \hline 
\multirow{6}{*}{ML-based ADM} 
& \multirow{3}{*}{UF} & 1620 & 12\\ 
\cline{3-4} & & 1500 & 10\\ 
\cline{3-4} & & 1380 & 8\\ 
\cline{2-4} 
& \multirow{3}{*}{OF} & 1260 & N/A\\
\cline{3-4} & & 1140 & 10\\ 
\cline{3-4} & & 1020 & 8
\\ \hline 
\end{tabular} 
\vspace{-6pt}
\end{table}
%%%%%%%%%%%%%%%%%%%%%%%%%%%%%

%%%%%%%%%%%%%%%%%%%%%%%%%%%%%
%%%%%%%%%%%%%%%%%%%%%%%%%%%%%
\subsection{Evaluation of Scalability of the Framework}
\label{subsec:eval_scalability}
We assess the scalability of the \framework~by examining how the time required for attack analysis changes with varying attack duration, defined as the number of time slots during which the attack occurs. This evaluation is illustrated in Table~\ref{tab:scalability}, which provides a comprehensive view of how the system's performance scales with different attack lengths. The table demonstrates a linear relationship between the time required for attack analysis and the duration of the attack. This indicates that as the size of the attack increases, the time needed for the system to analyze it also increases proportionally. However, the execution times are significantly longer than the LFC time. For instance, in the case of an attack spanning 1200 time slots (equivalent to 20 seconds), the analytics take between 5 to 14 minutes to identify the attack vector. Therefore, the analytics cannot be used for real-time attack vector identification. An attacker would need to be aware of the benign load pattern to execute such an attack or must time the attack to occur during periods when the load changes very minimally. This relationship is crucial for understanding the framework's scalability, as it highlights that the framework's performance is closely tied to the number of constraints (clauses) the solver must handle to synthesize the attack vector. To better understand these scalability issues, we present a detailed complexity analysis of the proposed analytics in Table~\ref{tab:complexity-analysis}.

%%%%%%%%%%%%%%%%%%%%%%%%%%%%%
\begin{table}[!t] 
\scriptsize 
\centering 
\label{tab:scalability}
\caption{Scalability Analysis of the System} 
\vspace{-6pt}
\begin{tabular}{|l|p{1.5cm}|p{2cm}|p{2cm}|} 
\hline 
Type of Defense & Attack Goal & \begin{tabular}[c]{@{}c@{}}Access to Timeslots\end{tabular} & Execution Time (second)
\\ \hline 
\multirow{6}{*}{\begin{tabular}[c]{@{}l@{}}Rules-Based\\ BDD\end{tabular}} 
& \multirow{3}{*}{UF} & 600 & 123.9 \\ 
\cline{3-4} & & 900 & 192.6 \\ 
\cline{3-4} & & 1200 & 283.8 \\ 
\cline{2-4} 
& \multirow{3}{*}{OF} & 600 & 132.2
\\ \cline{3-4} & & 900 & 202.2
\\ \cline{3-4} & & 1200 & 308.7\\ \hline 
\multirow{6}{*}{\begin{tabular}[c]{@{}l@{}}ML-based \\ ADM\end{tabular}} 
& \multirow{3}{*}{UF} & 600 & 234.5\\ 
\cline{3-4} & & 900 & 467.5\\ 
\cline{3-4} & & 1200 & 754.4\\ 
\cline{2-4} 
& \multirow{3}{*}{OF} & 600 & 298.4\\
\cline{3-4} & & 900 & 512.3\\ 
\cline{3-4} & & 1200 & 863.4
\\ \hline 
\end{tabular} 
\vspace{-6pt}
\end{table}
%%%%%%%%%%%%%%%%%%%%%%%%%%%%%

%%%%%%%%%%%%%%%%%%%%%%%%%%%%%
\begin{table}[!t]
\scriptsize
\centering
\label{tab:complexity-analysis}
\caption{Complexity Analysis of \framework}
\vspace{-6pt}
\begin{tabular}{|p{2cm}|p{1.5cm}|p{1.5cm}|p{2cm}|}
\hline
Phase                          & Time Complexity                     & Space Complexity   & Notation Description    \\ \hline
ADMs' Cluster Formation        & $O(|D|^2)$                          & $O(|D|)$           & $D$: All data samples $|D|$: Number of data samples for model training \\ \hline
Attack Constraints Formulation & $O(|\mathcal{S}|)$ & $O(|\mathcal{S}|)$ & $|\mathcal{S}|$ Number of sensor measurements  \\ \hline
Threat Vector Extraction       & $O(2^c)$                            & $O(|\mathcal{S}|)$ & $c$: Number of clauses    \\ \hline
\end{tabular}
\vspace{-6pt}
\end{table}
%%%%%%%%%%%%%%%%%%%%%%%%%%%%%
%%%%%%%%%%%%%%%%%%%%%%%%%%%%%
%%%%%%%%%%%%%%%%%%%%%%%%%%%%%%%%%%%%%%%%%%%%%
%%%%%%%%%%%%%%%%%%%%%%%%%%%%%%%%%%%%%%%%%%%%%
\section{Related Work}
\label{sec:related-work}

Modern power grids face a cybersecurity risk as enhanced monitoring and anomaly detection become more widely used. Malicious data can be transmitted to the control center by adversaries using compromised measuring devices, which could result in physical harm. For the safety of power systems, cyberattack scenario analysis is essential. Studies on FDI attacks assume a full or partial understanding of the targeted system. Recent research, such as \cite{ref_a, ref_b, ref_c}, focuses on manipulating load and power transmission measures with fake data. With a limited understanding of the power system, Chen et al.~\cite{ref_fdi_agc_rl} demonstrated remote FDI attacks on the power grid. 
%Sou et el. has proposed a Fuzzy C-means algorithm to identify anomaly detection in real time~\cite{Sou2023}. Another two data-driven ML-based attack detection and identification approaches by load altering in power systems have been demonstrated in~\cite{Sthapit2022}.

An FDI attack aims to skew measurements to influence load frequency control, var voltage control, and automated voltage regulation \cite{ref_x}. Sensor data is sent to control centers in contemporary power systems so that optimization and control algorithms can analyze it. Lou et al. The communication channel that transfers data from the power sensor to the control centers is the target of the FDI attack in this article\cite{ref_kaur}. Power system operators and researchers utilize advanced techniques and instruments to detect and reduce inaccurate or compromised data, which may pose a risk to cybersecurity or cause system instability or operational problems. However, if the attack is sufficiently covert, the attacker can get behind the control center's BDD or other anomaly detection model, causing the system to become unstable due to malicious input.

Analyzing and addressing constraints in ML-based systems pose distinct challenges compared to rule-based systems. The formal analysis of deep neural network (DNN)-based ML models has garnered attention in contemporary research. Various tools like Reluplex, Sherlock, and Marabou have been developed for verifying DNNs~\cite{katz2017reluplex, dutta2017output, katz2019marabou}. Dreossi et al. explored issues arising when applying formal methods to ML and analyzed ML-based system behavior in the presence of environmental uncertainty~\cite{dreossi2019verifai}. %T$\ddot{o}$rnblom et al. presented a tool called VoTE (Verifier of Tree Ensembles) for verifying DT-based ensemble techniques supporting up to 25 trees. %Souri et al. formally verified a hybrid ML-based approach for fault prediction in IoT applications, incorporating Multi-layer Perceptron (MLP) and Particle Swarm Optimization (PSO) algorithms~\cite{souri2020formal}. 
In contrast to these efforts, our proposed framework conducts RL-based threat analysis of smart home control systems, evading ML-based time-series ADM for synthesizing attacks, paving the way for a new research direction for the RL domain.

Formal analysis efficiently synthesizes verifiable attack vectors; however, existing efforts overlook the integration of ML-based IoT-enabled CPSs. Adversarial ML and RL-based techniques are efficient and fast, yet they often fail to extract attacks that can evade ML-based ADMs. Their primary limitation is the lack of guarantees for optimal or verifiable attack vector identification, stemming from their non-adherence to established grid and ADM constraints, unlike formal methods. Some of our notable research works, such as SHATTER~\cite{haque2023shatter} and SHChecker~\cite{haque2021novel}, identify optimal FDI attacks to provide wrong control signal to the corresponding safety-critical CPS bypassing different ML models~\cite{jafari2022optimal}. SHATTER effectively identifies attack vectors in ML-based smart home systems by focusing on time-series anomaly detection models (ADMs) and accounting for multi-time-slot attacks. Identifying the attack vector is similar to solving a scheduling problem, which is NP-hard. To manage this complexity, the problem space is divided into smaller time windows. However, SHATTER's analysis is constrained by its reliance on simple controller dynamics. In contrast, SHChecker improves upon this by modeling the controller or decision model using a non-linear machine learning approach. Nevertheless, it works for considered ADMs, and attacks are applicable for single-time instances. Jafari et al. identified FDI attacks by adopting a distinct approach to exploring the dynamics of complex power systems. Yet, their research predominantly depends on a rules-based BDD. Ni et al. introduced and assessed a new reinforcement learning method designed to determine the minimum number of attacks or actions required to trigger a blackout in smart grid systems, even when the attacker possesses only limited topological information about the power network \cite{ni2017reinforcement}. %Gohil et al. developed an automated, scalable, and practical framework utilizing reinforcement learning called Attrition. This framework effectively evades several advanced hardware Trojan (HT) detection methods, including logic testing and side-channel analysis~\cite{gohil2022attrition}. 
Nevertheless, the scope of these solutions is restricted to the generation of attacks in real time.

Jafari et al. propose an optimization-based formal model for FDI attacks on power systems, specifically targeting frequency-based protection relays. The model aims to find the optimal FDI attacks with the minimum required time to trigger a false relay operation, considering the dynamic behavior of the power system. The probability of attack success has been analyzed based on various system properties~\cite{jafari2023optimal}. \framework~has several distiction with the TIFS'23 work. Firstly, \framework~performs impact-based security analysis based on security properties such as resiliency, ADM robustness, adversarial capability, etc.
Conversely, TIFS'23 focuses on system properties such as governor inertia and droop to analyze the system's security. Secondly, the BDD considered in the TIFS'23 work is straightforward, and most of the attack vectors identified from the framework are detectable by ML-based ADMs. For instance, the DBSCAN-based ADM considered in our work detected 100\% of attacks from earlier research. Thirdly, the TIFS'23 work assumes that the attack goal will be successful by maliciously triggering all relays (load shedding, over-frequency, or ROCOF relays). However, the dynamics will change once a relay is triggered and isolates loads or generators. The TIFS'23 did not consider this in the modeling. Our work addresses this by limiting our attack goal to tripping a single load or generator. 

Beyond the mentioned literature, some works focus on attack detection through network packet analysis using ML-based  approaches~\cite{rawat2022modeling}. Hinta et al. outlined vulnerabilities in the Message Queue Telemetry Transport (MQTT) protocol, a critical IoT protocol~\cite{hintaw2021mqtt}. Our framework does not identify vulnerabilities in IoT protocols; instead, it pinpoints FDI attack vectors and potential impacts from an exploited IoT system. %Ramaki et al. developed a framework for detecting advanced persistent threats in critical infrastructure by modeling graph-based attackers' behavior~\cite{ramaki2023captain}. 
While these approaches identify attacks based on learned threat patterns, our framework synthesizes attack paths bypassing an ML-based ADM using formal analysis.
%%%%%%%%%%%%%%%%%%%%%%%%%%%%%
%%%%%%%%%%%%%%%%%%%%%%%%%%%%%%%%%%%%%%%%%%%%%
%%%%%%%%%%%%%%%%%%%%%%%%%%%%%%%%%%%%%%%%%%%%%
\section{Discussion and Limitations}
\label{sec:discussions}

\framework~outlined a framework for identifying attack vectors from ML-based ADM-assisted safety-critical CPSs. Unlike the previous works, our proposed attack analyzer can identify attack vectors from multi-timeslot-based complex dynamics. The findings of our work can be summarized as follows.

\begin{itemize}
    \item The successful completion of an attack requires more time for the SG with ML-based ADM than BDD-based rules. Hence, the ADM provides a certain level of resiliency against attacks.
    \item The linearization process loses certain details of the ADM. Hence, we can observe some deviation between the estimated attack impact from \framework~and the impact verified by OPALRT.
    \item The time for identifying the attack vector grows linearly (approximately degree 2) with the increase in the possible attack duration.
\end{itemize}

%\noindent \textbf{Possible Future Extensions:} ~
%\framework~has opened up research directions for a wide group of possible extensions. 
We have identified the following limitations of our proposed framework. 
%to plan for the future extension. 

\begin{itemize}
    \item \textbf{System Robustness:} The proposed framework is an efficient tool for the security and resiliency analysis of the SG system. Based on the identified attack vectors, \framework~can provide a defense guide. However, our proposed framework doesn't develop a more resilient defense tool using the attack vectors and corresponding impact. %The work will be further extended with an automated CPS reconfiguration scheme leveraging the attack vectors identified by the framework.
    \item \textbf{Analyzing System with Practical ADM:} Although we model a generic framework, we consider the relationship between the current load measurement and the previous one for designing the ADMs. The attack impact can be significantly reduced by considering ADM with multiple lookbacks (i.e., the relationship between multiple load measurements). However, ADM with a large lookback value might not be practical since that will raise a lot of false alarms in the system. Moreover, the ADM can be further improved through learning the relationship between consecutive frequency measurements. %Our future work will analyze an optimal ADM for more practical attack vector identification. Additionally, MISGUIDE will be updated to support a wide group of ADMs, if not all. We will further explore identifying attack vector extraction from ensembled-based ML models.
    \item \textbf{Analyzing Cascaded Failures:} The current attack goal is limited to triggering the UF or OF relays. The actual attack impact can be realized after a substation load or generator has been disconnected from the system. The disconnecting of a generator can lead to cascading failure of the generators and result in destruction, which needs to be analyzed. %The framework will be updated to assess the potential for cascading attack impacts, aiming to enhance the effectiveness of the ADM to combat those attacks.
    \item \textbf{Large Scale System Analysis:} The current analysis is carried out on a 39-bus system since the available real-world load data was collected from the IEEE-39 bus system. %In the future, we will analyze the scalability of the analyzer on larger-scale systems.
\end{itemize}

We plan to address these limitations in future works.

%%%%%%%%%%%%%%%%%%%%%%%%%%%%%
%%%%%%%%%%%%%%%%%%%%%%%%%%%%%%%%%%%%%%%%%%%%%
%%%%%%%%%%%%%%%%%%%%%%%%%%%%%%%%%%%%%%%%%%%%%
\section{Conclusion}
\label{sec:conclusion}

In this work, we proposed a novel defense-aware framework, \framework, to identify stealthy, verifiable, and multi-timeslot-based attack vectors while considering the complex SG dynamics. We validate our proposed framework on the IEEE 39-bus system with actual load data. Our experiments demonstrate that attack vectors synthesized by \framework~falsely trip the protective relays in minimal time (optimal) while evading the ML-based ADMs (stealthy). For our ADM, we consider the relationship between the current load measurement and the one immediately before it. While analyzing multiple previous load measurements significantly reduces the attack space, it can also increase false positives. Hence, we will address this in future research by determining the optimal number of previous measurements (i.e., $l$) required.
Furthermore, we will also compare the performances of various ADMs other than DBSCAN in identifying the attack vectors. We consider large-scale systems for scalability analysis and conduct a hardware-in-the-loop simulation using OPALRT to verify the applicability of the attack vectors. We replicate the transient behavior by considering the 10-minute separated load measurements of the GEFCom dataset as consecutive measurements. We will collaborate with the power grid facilities for a more realistic system modeling, verify the synthesized attack vectors, and analyze the system's resiliency.
%%%%%%%%%%%%%%%%%%%%%%%%%%%%
\bibliographystyle{unsrt}
\bibliography{References, bibliographies/LFC, bibliographies/Introduction, bibliographies/Problem-Definition, bibliographies/Related-Work} 

\begin{thebibliography}{10}

\bibitem{yu2016sg}
Xinghuo Yu and Yusheng Xue.
\newblock Smart grids: A cyber–physical systems perspective.
\newblock {\em Proc. IEEE}, 104(5):1058--1070, 2016.

\bibitem{bose2010smart}
Anjan Bose.
\newblock Smart transmission grid applications and their supporting infrastructure.
\newblock {\em IEEE Transactions on Smart Grid}, 1(1):11--19, 2010.

\bibitem{gregory2023securityintelligence}
Jennifer Gregory.
\newblock Today’s biggest threats against the energy grid (2023).
\newblock 2023.
\newblock Accessed: 2024-03-21.

\bibitem{li2017cybersecurity}
Zhiyi Li, Mohammad Shahidehpour, and Farrokh Aminifar.
\newblock Cybersecurity in distributed power systems.
\newblock {\em Proc. IEEE}, 105(7):1367--1388, 2017.

\bibitem{liang2015}
Gaoqi Liang, Steven~R. Weller, Junhua Zhao, Fengji Luo, and Zhao~Yang Dong.
\newblock The 2015 ukraine blackout: Implications for false data injection attacks.
\newblock {\em IEEE Trans. Power Syst.}, 32(4):3317--3318, 2016.

\bibitem{bobba2010detecting}
Rakesh~B Bobba, Katherine~M Rogers, Qiyan Wang, Himanshu Khurana, Klara Nahrstedt, and Thomas~J Overbye.
\newblock Detecting false data injection attacks on {DC} state estimation.
\newblock In {\em Preprints of the First Workshop on Secure Control Sys., CPSWEEK}, volume 2010. Stockholm, Sweden, 2010.

\bibitem{liu2011false}
Yao Liu, Peng Ning, and Michael~K Reiter.
\newblock False data injection attacks against state estimation in electric power grids.
\newblock {\em ACM Trans. Inf. and Syst. Secur. (TISSEC)}, 14(1):1--33, 2011.

\bibitem{karnouskos2011stuxnet}
Stamatis Karnouskos.
\newblock Stuxnet worm impact on industrial cyber-physical system security.
\newblock In {\em Proc. IECON 37th Annu. Conf. of the IEEE Ind. Electron. Soc.}, pages 4490--4494, 2011.

\bibitem{Dragonfly}
Hackers infiltrated power grids, 2014.

\bibitem{rocof2020}
M.~{Grebla}, J.~R. A.~K. {Yellajosula}, and H.~K. {Høidalen}.
\newblock Adaptive frequency estimation method for {ROCOF} islanding detection relay.
\newblock {\em IEEE Trans. Power Delivery}, 35(4):1867--1875, 2020.

\bibitem{haque2021novel}
Nur~Imtiazul Haque, Mohammad~Ashiqur Rahman, Md~Hasan Shahriar, Alvi~Ataur Khalil, and Selcuk Uluagac.
\newblock A novel framework for threat analysis of machine learning-based smart healthcare systems.
\newblock {\em arXiv preprint arXiv:2103.03472}, 2021.

\bibitem{jafari2022optimal}
Mohamadsaleh Jafari, Mohammad~Ashiqur Rahman, and Sumit Paudyal.
\newblock Optimal false data injection attacks against power system frequency stability.
\newblock {\em IEEE Transactions on Smart Grid}, 14(2):1276--1288, 2022.

\bibitem{hong2016probabilistic}
Tao Hong, Pierre Pinson, Shu Fan, Hamidreza Zareipour, Alberto Troccoli, and Rob~J Hyndman.
\newblock Probabilistic energy forecasting: Global energy forecasting competition 2014 and beyond.
\newblock {\em International Journal of forecasting}, 32(3):896--913, 2016.

\bibitem{misguide2024}
Misguide.
\newblock \url{https://github.com/misguidetdsc/misguide}, 2024.

\bibitem{jafari2023optimal}
Mohamadsaleh Jafari, Mohammad~Ashiqur Rahman, and Sumit Paudyal.
\newblock Optimal false data injection attack against load-frequency control in power systems.
\newblock {\em IEEE Transactions on Information Forensics and Security}, 2023.

\bibitem{zhu2018tu}
Yanzi Zhu, Zhujun Xiao, Yuxin Chen, Zhijing Li, Max Liu, Ben~Y Zhao, and Haitao Zheng.
\newblock Et tu alexa? when commodity wifi devices turn into adversarial motion sensors.
\newblock {\em arXiv preprint arXiv:1810.10109}, 2018.

\bibitem{ding2018safety}
Wenbo Ding and Hongxin Hu.
\newblock On the safety of iot device physical interaction control.
\newblock In {\em Proceedings of the 2018 ACM SIGSAC Conference on Computer and Communications Security}, pages 832--846, 2018.

\bibitem{gurobilogicalconst2024}
Jaromił Najman.
\newblock How do i model conditional statements in gurobi?, 2024.
\newblock Accessed: 09/07/2024.

\bibitem{ref_a}
Yanling Yuan, Zuyi Li, and Kui Ren.
\newblock Modeling load redistribution attacks in power systems.
\newblock {\em IEEE Transactions on Smart Grid}, 2(2):382--390, 2011.

\bibitem{ref_b}
Yanling Yuan, Zuyi Li, and Kui Ren.
\newblock Quantitative analysis of load redistribution attacks in power systems.
\newblock {\em IEEE Transactions on Parallel and Distributed Systems}, 23(9):1731--1738, 2012.

\bibitem{ref_c}
Xuan Liu and Zuyi Li.
\newblock Local load redistribution attacks in power systems with incomplete network information.
\newblock {\em IEEE Transactions on Smart Grid}, 5(4):1665--1676, 2014.

\bibitem{ref_fdi_agc_rl}
Ying Chen, Shaowei Huang, Feng Liu, Zhisheng Wang, and Xinwei Sun.
\newblock Evaluation of reinforcement learning-based false data injection attack to automatic voltage control.
\newblock 10(2):2158--2169, 2019.

\bibitem{ref_x}
Siddharth Sridhar and Manimaran Govindarasu.
\newblock Model-based attack detection and mitigation for automatic generation control.
\newblock {\em IEEE Transactions on Smart Grid}, 5(2):580--591, 2014.

\bibitem{ref_kaur}
Rajvir Kaur, Justin Albrethsen, David~K.Y. Yau, and Shahram Ghahremani.
\newblock Vulnerability assessment of false data injection attacks on optimal power flow.
\newblock In {\em 2021 IEEE PES Innovative Smart Grid Technologies - Asia (ISGT Asia)}, pages 1--5, 2021.

\bibitem{katz2017reluplex}
G.~Katz, C.~Barrett, D.~Dill, K.~Julian, and M.~Kochenderfer.
\newblock Reluplex: An efficient smt solver for verifying deep neural networks.
\newblock In {\em International Conference on Computer Aided Verification}, pages 97--117. Springer, 2017.

\bibitem{dutta2017output}
S.~Dutta, S.~Jha, S.~Sanakaranarayanan, and A.~Tiwari.
\newblock Output range analysis for deep neural networks.
\newblock {\em arXiv preprint arXiv:1709.09130}, 2017.

\bibitem{katz2019marabou}
G.~Katz, D.~Huang, D.~Ibeling, K.~Julian, C.~Lazarus, R.~Lim, P.~Shah, S.~Thakoor, H.~Wu, A.~Zelji{\'c}, et~al.
\newblock The marabou framework for verification and analysis of deep neural networks.
\newblock In {\em International Conference on Computer Aided Verification}, pages 443--452. Springer, 2019.

\bibitem{dreossi2019verifai}
Tommaso Dreossi, Daniel~J Fremont, Shromona Ghosh, Edward Kim, Hadi Ravanbakhsh, Marcell Vazquez-Chanlatte, and Sanjit~A Seshia.
\newblock Verifai: A toolkit for the formal design and analysis of artificial intelligence-based systems.
\newblock In {\em International Conference on Computer Aided Verification}, pages 432--442. Springer, 2019.

\bibitem{haque2023shatter}
Nur~Imtiazul Haque, Maurice Ngouen, Mohammad~Ashiqur Rahman, Selcuk Uluagac, and Laurent Njilla.
\newblock Shatter: Control and defense-aware attack analytics for activity-driven smart home systems.
\newblock {\em arXiv preprint arXiv:2305.09669}, 2023.

\bibitem{ni2017reinforcement}
Z.~Ni, S.~Paul, X.~Zhong, and Q.~Wei.
\newblock A reinforcement learning approach for sequential decision-making process of attacks in smart grid.
\newblock In {\em 2017 IEEE Symposium Series on Computational Intelligence (SSCI)}, pages 1--8. IEEE, 2017.

\bibitem{rawat2022modeling}
Romil Rawat, Vinod Mahor, Bhagwati Garg, Mukesh Chouhan, Kiran Pachlasiya, and Shrikant Telang.
\newblock Modeling of cyber threat analysis and vulnerability in iot-based healthcare systems during covid.
\newblock In {\em Lessons from COVID-19}, pages 405--425. Elsevier, 2022.

\bibitem{hintaw2021mqtt}
Ahmed~J Hintaw, Selvakumar Manickam, Mohammed~Faiz Aboalmaaly, and Shankar Karuppayah.
\newblock Mqtt vulnerabilities, attack vectors and solutions in the internet of things (iot).
\newblock {\em IETE Journal of Research}, pages 1--30, 2021.

\end{thebibliography}

\vspace{-40pt}
%%%%%%%%%%%%%%%%%%%%%%%%%%%%%%%%%
%%%%%%%% NI Haque %%%%%%%%%%%%%%%
\begin{IEEEbiography}[{\includegraphics[width=1in,height=1.15in,clip]{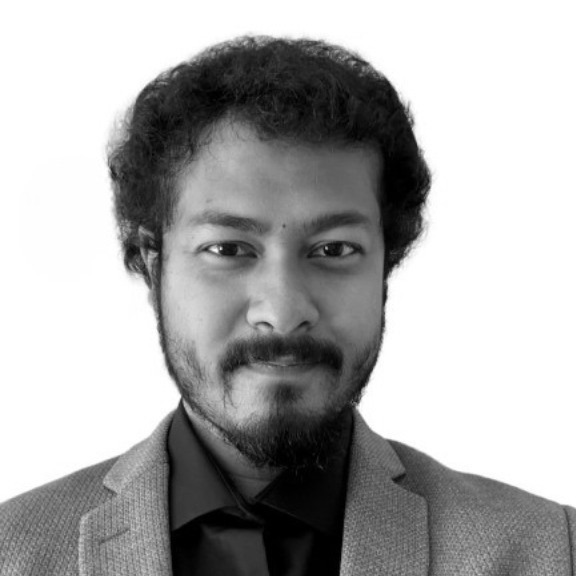}}]{Nur Imtiazul Haque} is a visiting assistant professor at University of Cincinnati. He completed his Ph.D. in Electrical and Computer Engineering (ECE) at Florida International University (FIU), Florida, USA .His research integrates formal methods, machine learning, and mathematical optimization to secure critical infrastructures against emerging cyber-physical system threats. During his Ph.D., he successfully contributed to and completed two NSF and one DOE projects, published 12 peer-reviewed papers and one book chapter, and patented a tool.
\end{IEEEbiography}
%%%%%%%%%%%%%%%%%%%%%%%%%%%%%%%%%
\vspace{-35pt}

%%%%%%%%%%%%%%%%%%%%%%%%%%%%%%%%%
%%%%%%%%%% Prabin Mali %%%%%%%%%%
\begin{IEEEbiography}[{\includegraphics[width=1in,height=1.25in,clip]{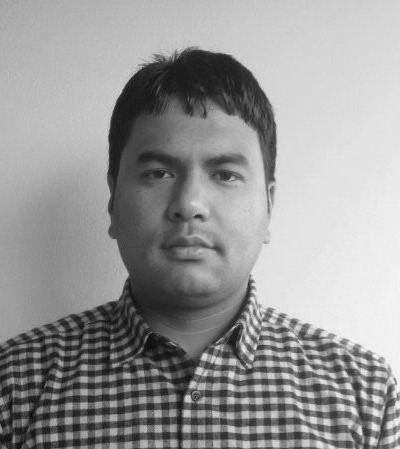}}]{Prabin Mali} is pursuing his Ph.D. degree in ECE at FIU, Miami, Florida, USA and has been working as a Graduate Research Assistant at Power System Computational Laboratory (PSCL). He has completed his Bachelor's Degree in Electrical Engineering and Master's Degree in Power System Engineering from Tribhuvan University (TU), Nepal. He is currently working in the sector of modeling of transmission and distribution systems with a keen interest in applying machine learning algorithms to power systems.
\end{IEEEbiography}
\vspace{-35pt}

%%%%%%%%%%%%%%%%%%%%%%%%%%%%%%%%%
%% Mohammad Zakaria Haider %%%%%%
\begin{IEEEbiography}[{\includegraphics[width=1in,height=1.25in,clip]{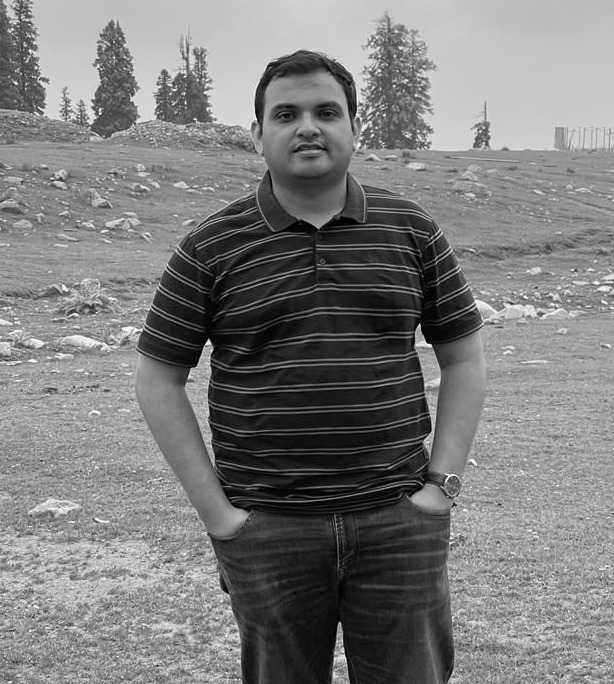}}]{Mohammad Zakaria Haider} completed his Bachelor of Science in Electrical and Electronic Engineering from Bangladesh University of Engineering and Technology (BUET) in March 2016. After his graduation, he worked different functions at Grameenphone, the leading telecom operator in Bangladesh, till August 2023. His field of research includes artificial intelligence with cyber-physical systems, focusing on strengthening security in the cyber domain.
\end{IEEEbiography}
\vspace{-40pt}

%%%%%%%% MA Rahman %%%%%%%%%%%%%%%
\begin{IEEEbiography}[{\includegraphics[width=1in,height=1.25in,clip]{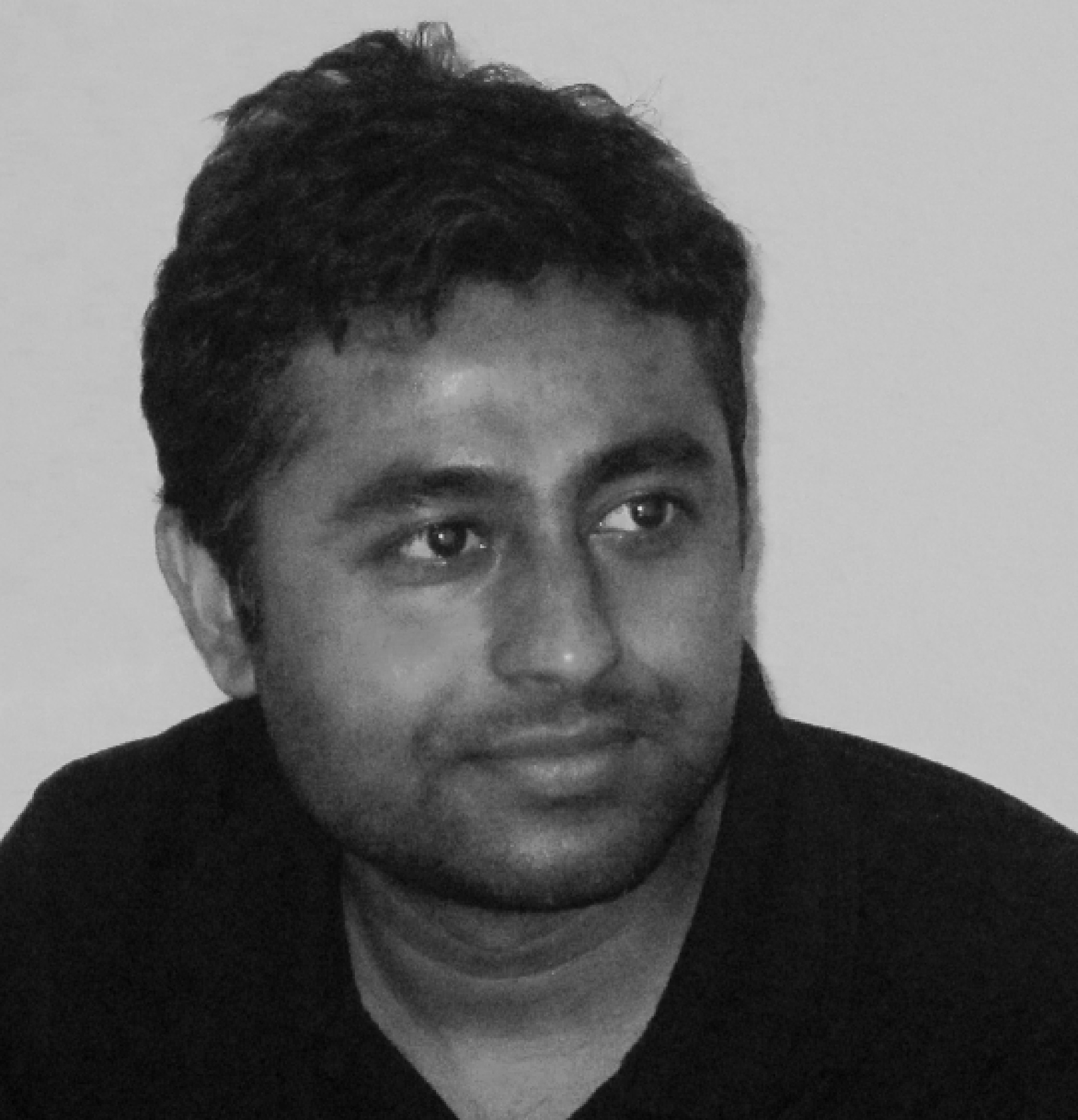}}]{Mohammad Ashiqur Rahman} is an Associate Professor in the Department of ECE at FIU, USA. He is leading the Analytics for Cyber Defense (ACyD) Lab at FIU. Before joining FIU, he was an Assistant Professor at Tennessee Tech University. He obtained the PhD degree in computing and information systems from the University of North Carolina at Charlotte in 2015. The focus of his research primarily includes artificial intelligence-based novel analytics design and development for network and information security, control-aware resiliency, and security hardening.
\vspace{-40pt}
\end{IEEEbiography}

%%%%%%%%%%%%%%%%%%%%%%%%%%%%%%%%%
%%%%%%%% Sumit Paudyal%%%%%%%%%%%
\begin{IEEEbiography}[{\includegraphics[width=1in,height=1.1in,clip]{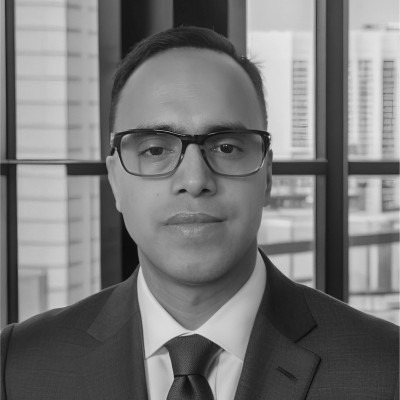}}]{Sumit Paudyal} is currently the Eminent Scholar Chaired Associate Professor in the Department of Electrical and Computer Engineering at FIU. He earned his PhD in Electrical Engineering from the University of Waterloo, Canada, in 2012, his MS in Electrical Engineering from the University of Saskatchewan, Canada, in 2008, and his BE in Electrical Engineering from Tribhuvan University, Nepal, in 2003. In 2018, Dr. Paudyal received the NSF CAREER Award and was named Best Professor of the Year by the Eta Kappa Nu (HKN) Student Society during his previous tenure at Michigan Tech. He currently serves as an Associate Editor for the IEEE Transactions on Smart Grid and the IEEE Transactions on Industry Applications.
\end{IEEEbiography}
\vspace{-10pt}

\end{document}